\documentstyle[12pt]{article}
\begin{document}
\title{Propagators in Curved Space}
\author{Frank Antonsen and Karsten Bormann\\
         Niels Bohr Institute\\
         Blegdamsvej 17\\
         DK-2100 Copenhagen \O}
\maketitle
\begin{abstract}
We demonstrate how to obtain explicitly the propagators for quantum
fields residing in curved space-time using the heat kernel for which
a new construction procedure exists. Propagators are determined for
the case of Rindler, Friedman-Robertson-Walker, Schwarzschild and
general conformally flat metrics, both for scalar, Dirac and Yang-Mills fields.
The calculations are based on an improved formula for the heat kernel in a
general curved space. All the calculations are done in $d=4$ dimensions for
concreteness, but are easily generalizable to arbitrary $d$. The new method
advocated here does not assume that the fields are massive, nor is it based
on an aymptotic expansion as such.\\
Whenever possible, the results are compared to that of other authors.
\end{abstract}

\section{Introduction}
The calculation of propagators of various fields in a curved background is
of utmost importance in theoretical physics, since it essentially provides
the only way of calculating scattering processes in, say, the early universe
or around a black hole. As there is at present no fully satisfactory theory 
of quantum gravity, the only way of studying quantum effects in the presence of strong
gravitational fields is through quantum field theory in curved spacetime, what
one might call semi-classical gravity.
The quantities of interest there are first and foremost the propagators, the
effective actions and the energy-momentum tensor. Therefore a lot of work have been
done in the past two or three decades in this field, see e.g. 
\cite{BD,Grib,Wald,Kay}.
In this paper we will concentrate
on the calculation of the propagators of quantum fields of spin zero, one-half 
and one. To the best of our knowledge this is the first time propagators of non-zero
spin fields is determined in a general background. We will leave the study of gravitons
on arbitrary backgrounds for future research -- the main aim of this paper is quantum
field theory in curved spacetime.\\
In previous papers \cite{Casimir,EffAct} we have developed 
a method for the determination of generating
functionals of quantum fields residing in curved space-time. This was
done using the heat kernel method and fairly reliable methods for
determining this quantity were devised \cite{Casimir}. 
One could decide to determine
propagators for the quantum fields from the generating functional 
but it turns out that one can construct
the propagators directly from the heat kernel and this latter approach is
the one pursued in this paper. It should be emphasized that the resulting
propagator is not just the free or bare one, as the calculations put forward
here are indeed non-perturbative, at least in the coupling to the background
fields (usually just the gravitational field, but occassional comments are 
made 
on how to include other kinds of backgrounds, e.g. external Yang-Mills or Higgs 
fields).\\
In section 2 we relate the heat kernel of a differential 
operator $A$, associated with
the quantum field in question, to the corresponding propagator. Then we
show how to explicitly determine propagators using previous results on
constructing the heat kernel. This is just the Schwinger-DeWitt
proper time formalism, but we use an improved expression for the heat kernel. 
Section 3 is devoted to a discussion of the corresponding vacuum
state and a discussion of the Hadamard condition.
We exemplify the approach in section 4 where we also compare our results to that
of other authors and finally
provide a brief conclusion and outlook in section 5.

\section{Determining the Propagator From the Heat Kernel}
Call the operator of interest, which generally varies in space, $A=A(x)$,
and the corresponding eigenfunctions $\psi_\lambda$ so that
\begin{equation}
    A\psi_\lambda=\lambda\psi_\lambda
\end{equation}
The heat kernel $G_A(x,x',\sigma)$ is the
function satisfying the heat kernel equation
\begin{displaymath}
    A G_A(x,x';\sigma) = -\frac{\partial}{\partial\sigma}G_A(x,x';\sigma)
    \label{eq:heat}
\end{displaymath}
subject to the boundary condition $G_A(x,x';0) =\delta(x,x')$. \\
Note that this equation is satisfied by (the spectral representation of the heat
kernel)
\begin{equation}
    G_A(x,x',\sigma)\equiv\sum_\lambda\psi_\lambda(x)\psi_\lambda^*(x')
    e^{-\lambda\cdot\sigma}\label{eq:heatseries}
\end{equation}
The Green's function on the other hand is a solution to the equation
\begin{equation}
    AG(x,x')=\delta(x,x')
\end{equation}
which is satisfied by (the spectral representation of the Green's function)
\begin{equation}
    G(x,x')\equiv\sum_\lambda\psi_\lambda(x)\psi_\lambda^*(x')\lambda^{-1}
\end{equation}
Put together with equation (\ref{eq:heatseries}) this yields the relationship 
between the Green's function and the heat kernel:
\begin{equation}
    G(x,x')=-\int_0^\infty d\sigma G_A(x,x';\sigma) \label{eq:Greens}
\end{equation}
(provided that $\lambda\neq 0$). This equation alone shows that the usual asymptotic
expansion, due to Schwinger and DeWitt, which is only valid for
$\sigma\rightarrow 0$, cannot be relied upon to provide good propagators. Furthermore,
for massless fields, the Schwinger-DeWitt expansion would lead to a manifestly 
divergent $\sigma$-integral. In the following section we will write down an
improved expansion for the heat kernel valid for $\sigma$ large as well, and which
moreover leads to closed expressions for the expansion coefficients. This improved
expansion is convergent even for $m=0$, and the method of calculation is readily
generalized to higher spin too. 

\subsection{Heat Kernel and Propagators for a Scalar Field}
The operator involved in the case of scalar particles is the (curved space)
d'Alembertian (minimally coupled scalar fields) or the d'Alembertian
plus some function (non-minimally coupled scalar fields). So we express
the d'Alembertian in  terms of the local vierbeins, $e^a_\mu\equiv
\frac{\partial x^a}{\partial x^\mu}$
where Greek indices refer to general coordinates while Latin
indices refer to the local inertial frame (the metric is $g_{\mu\nu}=
\eta_{ab}e^a_\mu e^b_\nu$ where $\eta_{ab}$ is the Minkowski metric,
$g$ is the determinant of the metric and $e$ the vierbein determinant, 
$e=\sqrt{g}$):
\begin{eqnarray}
    \Box&\equiv&\frac{1}{\sqrt{g}}\partial_\mu(\sqrt{g}g^{\mu\nu}\partial_\nu)
    \nonumber\\
  &=&\frac{1}{e}\partial_\mu(e\eta^{ab}e^\mu_ae^\nu_b\partial_\nu)\nonumber\\
  &=&\frac{1}{e}e^m_\mu\partial_m(e\eta^{ab}e^\mu_ae^\nu_be^n_\nu\partial_n)
    \nonumber\\
  &=&\frac{1}{e}e^m_\mu\partial_m(e\eta^{ab}e^\mu_ae^\nu_be^n_\nu)\partial_n
                  +\eta^{ab}e^m_\mu e^\mu_ae^\nu_be^n_\nu\partial_m\partial_n
    \nonumber\\
  &=&\Box_0+\frac{1}{e}e^m_\mu(\partial_m(ee^\mu_a))\partial^a
\end{eqnarray}
where $\Box_0=\eta^{ab}\partial_a\partial_b$ is the d'Alembertian of a 
comoving observer the heat kernel of which 
is known to be \cite{Grib}
\begin{equation}
    G_0(x,x';\sigma)=(4\pi\sigma)^{-2}e^{-\frac{\Delta(x,x')}{4\sigma}}
    \label{eq:flat}
\end{equation}
where $\Delta(x,x')\equiv(\int^x_{x'}ds)^2$, is the geodesic distance squared 
(i.e. half the so-called Synge world function) which is just $(x-x')^2$ in 
Cartesian coordinates.\footnote{The heat kernel 
(\ref{eq:flat}) is the straightforward covariant generalization of the result in
Minkowski space-time $G=(4\pi\sigma)^{-2}\exp(-\frac{(x-x')^2}{4\sigma})$. In
$d$ dimensions one simply has to replace $(4\pi\sigma)^{-2}$ by $(4\pi\sigma)
^{-d/2}$. The vierbeins are introduced to make this transition from flat to curved space
more transparent.} We will refer to $\Box_0$ as the ``flat'' d'Alembertian, and
consequently to $G_0$ as the ``flat'' heat kernel.
Proceed to remove the first order term of the Lagrangian by the substitution
\begin{equation}
G_\Box=\tilde{G}(x,x';\sigma)
        e^{-\frac{1}{2}\int\frac{1}{e}e^m_\mu(\partial_m(ee^\mu_n))dx^n}
        \label{eq:subst}
\end{equation}
The integral in this expression is always easy to calculate as the
term differentiated is just the reciprocal of the term with which it's
multiplied by so that one gets a logarithm of a product of vierbein 
components which in turn makes the exponential the reciprocal of
the squareroot of this product. It turns out, however, that we actually do not
need the explicit form of this integral as it cancels out in the final
expression.\\
When making the substitution (\ref{eq:subst}) in equation (\ref{eq:heat}) 
one gets a zero'th-order term
which must be added to the one that was there in the first place (i.e. to $\xi
R$).
One thus gets the following heat kernel equation ($\xi=0$ in the case of
minimal coupling):
\begin{equation}
\left[\Box_o+\frac{1}{4}\left(\frac{1}{e}e^m_\mu(\partial_m(ee^\mu_n))\right)^2
    -\frac{1}{2}\partial^n\left(\frac{1}{e}e_\mu^m\partial_m(ee_n^\mu)\right)+
    \xi R\right]\tilde{G}(x,x';\sigma)=-\partial_\sigma \tilde{G}(x,x';\sigma)
\end{equation}
written compactly as
\begin{equation}
      (\Box_0+f_0(x))\tilde{G}(x,x';\sigma)=-\partial_\sigma 
    \tilde{G}(x,x';\sigma)
\end{equation}
with
\begin{equation}
    f_0\equiv \xi R +\frac{1}{4}\left(e^{-1}\partial_\mu(ee^\mu_n)\right)^2 -
    \frac{1}{2}\partial^n\left(e^{-1}\partial_\mu(ee^\mu_n)\right) \label{eq:f0}
\end{equation}
The method for solving this equation devised in papers 
\cite{Casimir} makes use of the
following trick\footnote{One should note that this expansion is not the usual
asymptotic one due to Schwinger and DeWitt (see e.g. \cite{BD}), but 
actually holds for all values of
$\sigma$. Furthermore, as the recursion relation will show, it is actually
rather easy to evaluate the coefficients, especially along the diagonal $x=x'$.
The expansion shown here works for massless fields aswell, contrary to the
Schwinger-DeWitt expansion.}
\begin{equation}
    \tilde{G}(x,x';\sigma) = G_0(x,x';\sigma)e^{-T(x,x';\sigma)}
\end{equation}
Taylor expanding $T$
\begin{equation}
    T(x,x';\sigma) = \sum_{n=0}^\infty \tau_n(x,x')\sigma^n
\end{equation}
one gets the following recursion relation for the coefficients
\begin{equation}
    n\tau_n= -\Box_0\tau_{n-1}+\sum_{n'=0}^{n-1}\partial \tau_{n'}\cdot\partial
    \tau_{n-1-n'}+\frac{1}{2}\partial\Delta\cdot\partial\tau_n
\end{equation}
with $\tau_0=+\frac{1}{2}\int e^{-1}\partial_\mu(ee^\mu_n)dx^n$, as follows 
from the
boundary condition $\lim_{\sigma\rightarrow 0}G(x,x';\sigma) = \delta(x,x')$. 
Along the diagonal $x=x'$, this recursion relation simplifies
and the result is listed in \cite{Casimir}. 
In the general case $x\neq x'$, however, the
coefficients develop a dependency upon (the values of the curvature and its
derivatives along) the geodesic from $x$ to $x'$ (assuming
the existence of such a curve -- otherwise $\Delta$ has to be put equal to
$\infty$, leading to zero amplitude for the propagation). Explicitly,
\begin{equation}
    \tau_n(x,x') = e^{-2n\sqrt{\Delta(x,x')}}\int_0^1 g_{n-1}(\tau)
    e^{2n \tau}d\tau+c_ne^{-2n\sqrt{\Delta(x,x')}} \label{eq:closed}
\end{equation}
where the integral is along the abovementioned geodesic, and the functions
$g_{n-1}$ are given by
\begin{displaymath}
    g_n(x,x') = -\Box_0\tau_n(x,x')+\sum_{n'=1}^n \partial\tau_{n'} \cdot
    \partial\tau_{n-n'}
\end{displaymath}
with the first few functions being
\begin{eqnarray*}
    g_0 &=& \tilde{f}_0 \equiv f_0+\partial^n(e^{-1}\partial_\mu e^\mu_n) + 
    e^{-2}
    \eta^{mn}\partial_\mu(ee^\mu_m)\partial_\nu(ee^\nu_n)\\
    g_1 &=& -\Box_0\tau_1(x,x')+2\partial\tau_0\cdot\partial\tau_1\\
    g_2 &=& -\Box_0\tau_2(x,x')+(\partial\tau_1)^2+2\partial\tau_0\cdot
    \partial\tau_2
\end{eqnarray*}
and so on,
with $f_0$ given by (\ref{eq:f0}), $g_0$ is not just simply $f_0$, as one might
otherwise expect, a fact due to the
boundary condition $G(x,x';\sigma)\rightarrow \delta(x,x')$ for
$\sigma\rightarrow 0$. The variable $\tau$ is the proper-time of a comoving 
observer, scaled such that at $\tau=0$ the observer is at position $x$, whereas 
at the later time $\tau=1$ (s)he is at $x'$.\\ 
Inside the integral along the geodesic the differential operators $\partial, 
\Box_0$ simplify to become
\begin{eqnarray}
    \partial_p &=& e_p^\mu\partial_\mu = e_p^\mu\dot{x}_\mu(\tau)
    \frac{\partial}{\partial\tau} =
    \dot{x}_p(\tau)\frac{\partial}{\partial\tau} \label{eq:dif1}\\
    \Box_0 &=& \eta^{pq}\partial_p\partial_q = \eta^{pq}\dot{x}_p\ddot{x}_q
    \frac{\partial}{\partial\tau}+\dot{x}^2\frac{\partial^2}{\partial\tau^2}
        \label{eq:dif2}
\end{eqnarray}
where $x_\mu(\tau)$ describes the geodesic. These differentiations are
consequently rather straightforward to carry out.\\
The coefficients $c_n$ are independent of the
curve, they only depend upon the value at the end points, in fact 
\begin{equation}
    \lim_{x'\rightarrow x}\tau_n(x,x')=c_n(x)\mbox{   and   } \lim_{x\rightarrow
    x'}\tau_n(x,x')=c_n(x')
\end{equation}
and they are thus equal to the coefficients found in 
\cite{Casimir}. They satisfy
\begin{equation}
    (n+1)c_{n+1} = -\Box_0 c_n+\sum_{n'=0}^n \partial c_{n'}\cdot \partial 
    c_{n-n'}
\end{equation}
and thus the relevant ones are\footnote{Strictly speaking the coefficients are
$\frac{1}{2}(c_n(x)+c_n(x'))$, but for simplicity we have decided not to write
them in this way, thus a symmetrization in $x,x'$ is always understood. In the
particular examples this is carried out explicitly.}
\begin{eqnarray*}
    c_1 &=& \tilde{f}_0\\
    c_2 &=& -\frac{1}{2}\Box_0 f_0\\
    c_3 &=& \frac{1}{6}\Box_0^2 f_0 -\frac{1}{3}(\partial f_0)^2\\
        &\approx & -\frac{1}{3}(\partial f_0)^2
\end{eqnarray*}
where we have decided to only include first and second derivatives of $f_0$
(essentially the curvature) in the last approximation. These expressions are for
massless fields, masses are accomodated by adding $m^2$ to $f_0$, and therefore
to $\tau_1,c_1$.
Written out in full, the first non-trivial coefficient
function, $g_1$, is then
\begin{eqnarray*}
    g_1 &=& -\left[\frac{1}{2}\Delta^{-3/2}(\partial\Delta)^2 
        +\Delta^{-1/2}(\partial\Delta)^2-\Delta^{-1/2}\Box_0\Delta\right]\tau_1+
        \\
        && \Delta^{-1/2}(\partial^p\Delta)\left[\int_0^1\partial_p
            (g_0e^{2\tau})d\tau
        +\frac{1}{2}\partial_pc_1\right]e^{-2\sqrt{\Delta}}-\\
        && e^{-2\sqrt{\Delta}}\left[\int_0^1\Box_0(g_0 e^{2\tau})
        d\tau +\frac{1}{2}\Box_0c_1\right]+\\
        &&\eta^{mn}e^{-1}\partial_\mu(ee^\mu_n)\left[-\Delta^{-1/2}
            (\partial_m\Delta)
        \tau_1+e^{-2\sqrt{\Delta}}\left\{\int_0^1\partial_m(\tilde{f}_0
            e^{2\tau})d\tau+\partial_mc_1\right\}\right]
\end{eqnarray*}
where all derivatives outside the integrals are with respect to $x$ and not 
$x'$, the derivatives
inside the integrals, however, are to be understood as in 
(\ref{eq:dif1}-\ref{eq:dif2}). The next coefficient function, $g_2$, becomes
\begin{eqnarray*}
    g_2 &=& -\left[\Delta^{-3/2}(\partial\Delta)
    ^2+4\Delta^{-1/2}(\partial\Delta)^2-2\Delta^{-1/2}\Box_0\Delta\right] \tau_2
    +\\
    && 4\Delta^{-1/2}(\partial^p\Delta)\left[\int_0^1\partial_p(g_1e^{4\tau})
    d\tau+\frac{1}{2}\partial_pc_2\right] e^{-4\sqrt{\Delta}}-\\
    && e^{-4\sqrt{\Delta}}\left[\int_0^1\Box_0(g_1e^{4\tau})d\tau
    +\frac{1}{2}\Box_0c_2\right]+\\
    &&\left\{\Delta^{-1}(\partial\Delta)^2\tau_1^2+e^{-4\sqrt{\Delta}}
    \left[\int_0^1\partial_p(g_0e^{2\tau})d\tau+\frac{1}{2}\partial_pc_1\right]
    ^2\right.-\\
    &&\left.2\Delta^{-1/2}\tau_1e^{-2\sqrt{\Delta}}(\partial^p\Delta)\left[ 
    \int_0^1
    \partial_p(g_0e^{2\tau})d\tau+\frac{1}{2}\partial_pc_1\right]\right\}+\\
    &&\eta^{mn}e^{-1}\partial_\mu(ee^\mu_n)\left[-2\Delta^{-1/2}
    (\partial_m\Delta)\tau_2 +
    e^{-4\sqrt{\Delta}}\left\{\int_0^1\partial_m(g_1e^{4\tau})d\tau+\partial_m
    c_2\right\}\right]
\end{eqnarray*}
As is apparent, the
expressions quickly get somewhat involved. Note, however, that most of the
needed operations are trivial -- most of the terms are simple derivatives of
$\Delta$ and $\tilde{f}_0$. The only non-trivial part being the integral along
the geodesic. By the same token, we will usually only list $\tilde{f}_0$ and
$\tau_1$ in the examples to follow.\\
One should notice that in principle one could find the coefficients to any
chosen order, we will however stick to the approximate solution with $\tau_n
\approx 0~,~~ n\geq 4$, corresponding to $\partial^3 R, \partial^4 R,...$ 
being negligible. In fact we can make do with the even more 
tractable form
(which is adequate\footnote{As general relativity is only
renormalizable upto one loop order, it would not be meaningful to go to
higher order in this approximation; the higher order terms are essentially third
and higher order derivatives and powers of the curvature, corresponding to a 
higher loop order. This approximation corresponds to a regime in which $R\gg 
(\partial R)^2,\Box_0 R$, i.e. strong but slowly varying curvature. The other
extreme ($R\ll \partial R$, i.e. strongly varying curvatures, e.g. in the 
space-time
foam) can be treated similarly, the coefficients $c_n$ will then be $c_n \approx
(-1)^{n-1}\frac{1}{n!}\Box_0^{n-1}f_0$, and similarly for $g_n,\tau_n$.})
\begin{eqnarray}
      \tilde{G}(x,x';\sigma)&\approx& G_0(x,x';\sigma)e^{-\tau_1\sigma}(1+
      \tau_2(x,x')\sigma^2+\tau_3(x,x')\sigma^3)\nonumber\\
    &=& (4\pi\sigma)^{-2}e^{-\frac{\Delta(x,x')}{4\sigma}-\tau_1(x,x')
    \sigma}(1+\tau_2(x,x')\sigma^2+\tau_3(x,x')\sigma^3)
\end{eqnarray}
where expression (\ref{eq:flat}) for the ``flat'' space heat kernel has been 
inserted. \\
Other modifications of the Schwinger-DeWitt asymptotic expansion exists. In fact 
Parker and coworkers, \cite{Parker}, have suggested that a factor $\exp(-\xi R\sigma)$
was present if one performed a partial summation of the usual expansion. The basic
new thing in our approach is the actual possibility of a systematic calculation of
{\em all} the coefficients in the expansion -- in fact we've got a general, closed
expression, (\ref{eq:closed}). The expansion of Parker {\em et al.} differs
from ours in that whereas we expand $T$ in $\exp(T)$, they expand $\exp(T)$, 
thus getting simply modifications of the Schwinger-DeWitt coeffcients. The 
main point of the approach put forward in this paper is the actual abbility to
systematically find the expansion coefficients both for $x\neq x'$ (the 
general case) and for $x=x'$ (which turns out to be very simple). It is the
use of the exponential which makes this simplification possible.\\ 
As pointed out by Ford and Toms, \cite{FT}, there is a genuine need
for non-local terms in the heat kernel (and hence a need to go beyond the
Schwinger-DeWitt expansion) and as it turns out, the coefficients
obtained by the method put forward here are in fact non-local. Their geometric
meaning is even rather transparent as they involve the integration along a 
geodesic of various curvature-related quantities.\\
The heat kernel can also be written as a 
path-integral, \cite{path}, and the usual Van Vleck-Morette determinant occurs
when one considers geodesics, \cite{geod}.\\
The $\sigma$-integrals involved in determining the Green's function from
equation (5) are now of standard type leading to the following result \cite{GR}
\begin{eqnarray}
    G(x,x')\approx&-&(4\pi)^{-2}\left(2\left(\frac{\Delta(x,x')}{4\tau_1(x,x')}
    \right)^{-\frac{1}{2}}
    K_{-1}(\sqrt{\Delta(x,x')\tau_1(x,x')})+\right.\nonumber\\
    &&\tau_2(x,x')
    \left(\frac{\Delta(x,x')}{4\tau_1(x,x')}\right)^{\frac{1}{2}}
    K_1(\sqrt{\Delta(x,x')\tau_1(x,x')})-\nonumber\\
    &&\left.\tau_3(x,x')\frac{\Delta(x,x')}{4\tau_1(x,x')}K_{2}(
      \sqrt{\Delta(x,x')\tau_1(x,x')})\right) \label{eq:scalprop}
\end{eqnarray}
where $K_n$ is a modified Bessel (or MacDonald) function which has the 
following series 
expansion (needed in the case of gauge bosons treated below and 
for the proof of the Hadamard condition being satisfied):
\begin{eqnarray}
    K_n(x)&=& (-1)^{n+1}\sum_{k=0}^\infty\frac{1}{k!(n+k)!}\left(\ln\frac{1}{2}x
    -\frac{1}{2}\psi(n+k+1)-\frac{1}{2}\psi(k+1)\right)\left(\frac{x}{2}
    \right)^{2k+n}\nonumber\\
    &&\hspace{30mm}+\frac{1}{2}\sum_{k=0}^{n-1}(-1)^k\frac{(n-k-1)!}{k!} 
    \left(\frac{x}{2}\right)^{2k-n}
    \label{eq:Bessel}
\end{eqnarray}
with $\psi(x)=\frac{d}{dx}\ln\Gamma(x)$ is the Euler psi (or digamma) 
function.\\
Using the recursion relation \cite{GR}
\begin{displaymath}
    zK_{n-1}(z)-zK_{n+1}(z) = -2nK_n(z)
\end{displaymath}
together with $K_{-n}=K_n$, one can reexpress (\ref{eq:scalprop}) in terms of 
$K_0$ and $K_1=-\frac{d}{dz} K_0$, \cite{GR}, only as
\begin{equation}
    G(x,x') = \left[-\alpha_0(x,x')\frac{d}{dz}+\beta_0(x,x')\right]
    K_0(z) = \alpha_0(x,x')K_1(z)+\beta_0(x,x')K_0(z) \label{eq:exact}
\end{equation}
with $z=\sqrt{\Delta(x,x')\tau_1(x,x')}$. To the chosen order we have
\begin{eqnarray}
    \alpha_0(x,x') &=& -2(4\pi)^{-2}\left(\sqrt{\frac{4\tau_1}{\Delta}} + \tau_2
    \sqrt{\frac{\Delta}{4\tau_1}}+\tau_3\sqrt{\frac{\Delta}{\tau_1^3}}\right)
    \label{eq:alpha}\\
    \beta_0(x,x') &=& -\frac{\tau_3\Delta}{2\tau_1}(4\pi)^{-2} \label{eq:beta}
\end{eqnarray}
Note that the form (\ref{eq:exact}) of the Green's function is exact, it is only
the explicit expressions (\ref{eq:alpha}-\ref{eq:beta}) for $\alpha_0,
\beta_0$ which are approximate.\\
Thus the problem of finding the Green's function of a scalar field in some 
geometry, is reduced to evaluating the two functions $f_0,\tilde{f}_0$ (and from
these $\tau_n$),  
containing the information about the metric structure. The calculation of 
these is, as 
will also be apparent from the examples to follow, rather straightforward.\\
We will see that also the propagators for fields with non-zero spin can be
written in this form, involving only the two modified Bessel functions $K_0,
K_1$. The coefficients will contain all the spin-specific data, and will be
denoted by $\alpha_s,\beta_s$ for spin $s$. Only for $s=0$ are these scalar
functions, for bosons of spin $s$ they will in general be $2s$-tensors, whereas
for spinors they will take values in the Clifford algebra (for Rarita-Schwinger
fields, $s=3/2$, they would carry two Lorentz indices besides their two Clifford
algebra (i.e. spinor) ones, and so on).

\subsection{Heat Kernel and Propagators for Spin 1 Gauge Bosons}
The heat kernel for spin 1 gauge bosons has been determined in 
\cite{Casimir} and
because it is the result of a lengthy calculation we simply quote the
result.
The heat kernel equation can, in the mean field approximation chosen in 
\cite{Casimir}, be written 
as\footnote{In \cite{EffAct} we have calculated the effective action using 
the heat kernel without using a mean field approximation, but the result 
from \cite{Casimir}, 
which we use here, is conceptually simpler.}
\begin{eqnarray}
    &&\hspace{-5mm}\left[\delta^a_b\delta^m_n\partial_p\partial^p +\delta^a_b
      \left(\partial_n
    e^{m\mu}-\partial^me_n^\mu\right)e^p_\mu\partial_p+
    \langle gf_{b\hspace{3pt}c}^{\hspace{3pt}a}(\partial_nA^{mc}-\partial^m
    A^c_n)\rangle\right.\nonumber\\
    &&+\left.\langle\frac{1}{2}\delta^m_ng^2f_{ebc}f_d^{\hspace{3pt}ac}A^e_p
      A^{pd}\rangle\right] G_m^q(x,x';\sigma) =
    -\delta^a_b\frac{\partial G_n^q}{\partial\sigma}\label{eq:heat1}
\end{eqnarray}
where $G=G_m^q$ is a matrix valued function. Defining
\begin{equation}
    {\cal E}_n^{mp} = \left(\partial_ne^{m\mu}-\partial^me_n^\mu\right) e^p_\mu
\end{equation}
one can remove the first order term in equation (\ref{eq:heat1}) by substituting
(the quantities in the below equation are matrices)
\begin{equation}
    G = \tilde{G} e^{-\frac{1}{2}E}
\end{equation}
where $\partial^pE^m_n={\cal E}^{mp}_n$. The heat equation
then becomes
\begin{eqnarray}
    &&\delta^m_n\partial_p\partial^p\tilde{G}_m^q
    -\left[\frac{1}{2}\partial_p{\cal E}^{mp}_n+\frac{1}{4}{\cal E}_k^{mp}
    {\cal E}_{np}^k+\langle gf_{b\hspace{3pt}c}^{\hspace{3pt}a}(\partial_n
        A^{mc}-\partial^mA^c_n)\rangle+\right.\nonumber\\
    &&\hspace{20mm}\left.\langle\frac{1}{2}\delta^m_ng^2f_{ebc}
      f_{d\hspace{3pt}c}^{
      \hspace{3pt}a}A^e_pA^{pd}\rangle\right]
    \tilde{G}_m^q= -\frac{\partial}{\partial\sigma}\tilde{G}_n^q
\end{eqnarray}
an equation of the same functional form as for the
non-minimally coupled scalar field. The solution is then 
(in matrix notation)
\begin{equation}
    G(x,x';\sigma) = G_{\Box_0}(x,x';\sigma)e^{-{\cal A}\sigma
    +\frac{1}{2}{\cal B}\sigma^2-\frac{1}{3}{\cal C}\sigma^3}
\end{equation}
with\footnote{Even though the matrices ${\cal
A,B,C}$ do not in general commute, there is no problem with this solution, as
their commutator will inevitably be a higher order term. But the formula is 
not as useful as for the scalar case due to the
complications of the gauge field. On the other hand, the couplings and mean 
field values need not be small, as the result presented here is 
non-perturbative.}
\begin{eqnarray}
    {\cal A}_n^m &=& e^{-2\sqrt{\Delta}}\int_0^1 \tilde{\cal
    A}^m_n(x(\tau))e^{2\tau}d\tau +\tilde{\cal A}_n^me^{-2\sqrt{\Delta}} 
      \label{eq:A}\\
    {\cal B}_n^m &=& -\frac{1}{2}e^{-4\sqrt{\Delta}}\int_0^1 
    \Box_0{\cal A}_n^m e^{4\tau}d\tau+\tilde{\cal B}^m_ne^{-4\sqrt{\Delta}} 
      \label{eq:B}\\
    {\cal C}_n^m &\approx& -\frac{1}{3}e^{-6\sqrt{\Delta}}\int_0^1
    (\partial_p{\cal A}_k^m)(\partial^p{\cal A}^k_n)e^{6\tau}d\tau+\tilde{\cal
    C}^m_ne^{-6\sqrt{\Delta}}\label{eq:C}
\end{eqnarray}
where
\begin{eqnarray}
    \tilde{\cal A}^m_n &=& \partial_p{\cal E}^{mp}_n+\frac{3}{4}{\cal E}^{mp}_k 
    {\cal E}_{np}^k+\langle gf_{b\hspace{3pt}c}^{\hspace{3pt}a}(\partial_n
    A^{mc}-\partial^mA^c_n)\rangle+\nonumber\\
    &&\qquad\langle\frac{1}{2}\delta^m_ng^2f_{ebc}f_{d\hspace{3pt}c}^{
      \hspace{3pt}a}A^e_pA^{pd}\rangle
    \label{eq:tildeA}\\
    \tilde{\cal B}_n^m &=& \Box_0{\cal A}_n^m \label{eq:tildeB}\\
    \tilde{\cal C}^m_n &\approx& (\partial_p {\cal A}^m_k)(\partial^p 
    {\cal A}^k_n)
      \label{eq:tildeC}
\end{eqnarray}
corresponding to $c_1=\tilde{f}_0,c_2,c_3$ respectively, and
\begin{equation}
    \Box_0 \equiv \partial_p\partial^p = \eta^{pq}\partial_p\partial_q =
    g^{\mu\nu}\partial_\mu\partial_\nu+\eta^{mn}e_m^\mu(\partial_\mu e_n^\nu)
    \partial_\nu
\end{equation}
as before,
the heat kernel of which is given by equation (\ref{eq:flat}). Analogously to
the scalar case, the quantity $\tilde{\cal A}$ is essentially a curvature term.
The approximation
involved in the evaluation of ${\cal C},\tilde{\cal C}$ above is, as for the
scalar case, the omission of a fourth order derivative in the curvature, namely
$\Box_0^2{\cal A}$.\\
A few comments
concerning the actual calculation of this quantity are in order. Now, if the
matrices $\cal A,B,C$ were simply diagonal, one could apply the Bessel functions
directly to the diagonal values, which would then become scalar boson
propagators. Even when $\cal A$ is diagonal (which holds for abelian fields), it
will not {\em a priori} commute with $\cal B,C$ and these can therefore not in
general be diagonalized, leading to some technical difficulties in the practical
application of this formula. In  practical calculations it will 
of course be tempting, and, for most purposes, good enough, to ignore the 
${\cal B}$ and ${\cal C}$ terms of the heat kernel (which is strictly speaking
only good enough for abelian or weak fields and couplings) or to resort to 
numerical calculation of these higher order terms. 
This is unsatisfactory (and, moreover, rather 
cumbersome), but we have not yet been able to come up with a better solution.\\
Referring back to equation (\ref{eq:Greens}) the propagator becomes (compare
with the scalar case of equation(\ref{eq:scalprop})) 
\begin{eqnarray}
      D^m_n(x,x')=&-&(4\pi)^{-2}\left(2(\frac{1}{4}\Delta)^{-1/2}
      ({\cal A}^{1/2})^m_k \left(K_{-1}(\sqrt{\Delta {\cal A}})\right)_n^k+
      \right.\nonumber\\
        &&
      \frac{1}{2}{\cal B}^m_k (\sqrt{\frac{1}{4}\Delta}({\cal A}^{-1/2})^k_l 
      \left(K_{1}(\sqrt{\Delta {\cal A}})\right)_n^l-\nonumber\\
      &&\left.\frac{1}{3}{\cal C}^m_k\frac{1}{4}\Delta({\cal A}^{-1})^k_l
      \left(K_{2}(\sqrt{\Delta {\cal A}}))\right)_n^l
      \right)
\end{eqnarray}
which we will write as (in matrix notation)
\begin{equation}
     D(x,x') \equiv  \alpha_1(x,x')K_1(\sqrt{\Delta\tau_1})+\beta_1(x,x')
    K_0(\sqrt{\Delta\tau_1})\label{eq:YMprop}
\end{equation}
To the chosen order
\begin{eqnarray}
    \alpha_1 &=& 2(4\pi)^{-2}\left((\frac{1}{4}\Delta)^{-1/2}{\cal A}^{1/2}
    +(\frac{1}{4}\Delta)^{1/2}{\cal B}{\cal A}^{-1/2}+\frac{1}{4}\Delta 
    {\cal C}{\cal A}^{-1}\right)\\
    \beta_1 &=& \frac{1}{2}\Delta (4\pi)^{-2}{\cal C}{\cal A}^{-1}
\end{eqnarray}
with the Lorentz indices suppressed.\\
In order to determine the Green's function for a Yang-Mills field, all we have 
to do is to evaluate the matrix
$\tilde{\cal A}$, and from that the matrices $\tilde{\cal B},\tilde{\cal C}$. 
Just like the functions $f_0,\tilde{f}_0$ for the scalar
case, these matrices only depend upon the vierbein (and the mean field, which
can be expressed in terms of the vierbein as in \cite{Casimir}), 
but unlike the spin zero 
case, their evaluation is somewhat more complicated (but still feasible).\\
Lastly we have to consider the case of fermions:

\section{Heat Kernel and Propagators for Dirac Fermions}
Consider the heat equation for a Dirac field in curved space-time
\begin{equation}
    (i/\hspace{-3mm}\nabla-m)G_{1/2}(x,x';\sigma) = -\frac{\partial}{\partial
    \sigma} G_{1/2}(x,x';\sigma) \label{eq:heat1/2}
\end{equation}
where\footnote{It would be very easy to include, say, a coupling to a Yang-Mills
field $ig A_m^k\gamma^mT_k$, e.g. using the mean-field derived in the previous
section, or a Yukawa coupling to a scalar or pseudo scalar field. The latter two
would just amount to the addition of a term to the mass, whereas the Yang-Mills
coupling would appear in an extra contribution to the function $F_a$ to be
introduced below. The Yang-Mills field, however, would lead to a non-diagonal
form of $/\hspace{-3mm}\nabla^2$, as it would now contain a term like 
$\sigma^{\mu\nu}
F_{\mu\nu}^kT_k$, i.e. a spin-magnetic field coupling. The introduction of a
background torsion would lead to a similar term as shown in 
\cite{Casimir}.}
\begin{equation}
    /\hspace{-3mm}\nabla \equiv e^\mu_a\gamma^a(\partial_\mu+\omega_\mu^{bc}
    \sigma_{bc})
\end{equation}
This equation can be rewritten as
\begin{equation}
    (i\gamma^a(\partial_a+F_a)+i\gamma_5\gamma^a \tilde{F}_a-m)G_{1/2} =
    -\frac{\partial}{\partial\sigma}G_{1/2} \label{eq:Dirac}
\end{equation}
where
\begin{eqnarray}
    F_a &=& 8ie_d^\mu\omega_\mu^{bc}\delta^d_b\eta_{ac}\\
    \tilde{F}_a &=& -4ie_d^\mu\omega_\mu^{bc}\varepsilon_{bc~a}^{~~d} 
      \label{eq:tildeF}
\end{eqnarray}
Here we have used that the $\sigma_{\mu\nu}$, the generator of 
$SO(1,3)$-transformations when acting upon Dirac spinors, is an element of 
the Clifford algebra, $\sigma_{mn}
=\frac{1}{4}i\left[\gamma_m,\gamma_n\right]$. The spin connection 
$\omega_\mu^{ab}$ can be written in terms of derivatives of the vierbeins.
The heat kernel for a Dirac spinor takes values in the Clifford algebra, and 
can thus be expanded on the sixteen dimensional basis $1,\gamma_5,\gamma_a,
\gamma_5\gamma_a,\sigma_{ab}$. Due to this complication we have not been able to
find an expression for the heat kernel for a Dirac fermion. We can still find
the propagator, however, by noting
\begin{equation}
    S(x,x') = (i/\hspace{-3mm}\nabla+m)G'(x,x')
\end{equation}
is an inverse of the Dirac operator provided $G'$ satisfies
\begin{equation}
    (i/\hspace{-3mm}\nabla-m)(i/\hspace{-3mm}\nabla+m)G'=(-/\hspace{-3mm}
    \nabla^2-m^2)G'= 1
\end{equation}
Noting that
\begin{equation}
    /\hspace{-3mm}\nabla^2 = (\Box+\xi_f R)\mbox{\bf 1}
\end{equation}
where {\bf 1} is the unit element in the Clifford algebra (i.e. in this 
particular
case a $4\times 4$ unit matrix) we see that $G'$ is the Green's function for a 
non-minimally coupled scalar field. 
Had we considered a spin-3/2 fermion (a Rarita-Schwinger field) we would have
made the same {\em Ansatz} but with $G'$ replaced by $D'$, a similar propagator
for a spin-1 boson.\\
The propagator then becomes 
\begin{eqnarray}
    S(x,x') &=& (i/\hspace{-3mm}\nabla+m)(\alpha_0K_1+\beta_0K_0)\nonumber\\
    &\equiv & \alpha_{1/2}K_1+\beta_{1/2}K_0
\end{eqnarray}
where $\alpha_{1/2},\beta_{1/2}$ are Clifford algebra-valued. To the chosen
approximation, these new coefficient functions are
\begin{eqnarray}
    \alpha_{1/2} &=& (i/\hspace{-3mm}\nabla+m)\alpha_0-i(z^{-1}\alpha_0+\beta_0)
    /\hspace{-3mm}\nabla z\\
    \beta_{1/2} &=& (i/\hspace{-3mm}\nabla+m)
    \beta_0-i\alpha_0/\hspace{-3mm}\nabla z
\end{eqnarray}
with
\begin{equation}
    z\equiv\sqrt{\Delta\tau_1}
\end{equation}
We have now written $S(x,x')$, the propagator for a Dirac field, on the same
form as that of $G$, the scalar field propagator.

\subsection{On the Corresponding Vacuum and the Hadamard Condition}
Now, in order for a two point function to be a propagator, it must be possible
for it to be written as the expectation value of the (path ordered) product of
two field operators with respect to some vacua $\langle 0_{\rm out}|, |0_{\rm
in}\rangle$. As is well-known, \cite{BD,Grib,Wald}, the concept of a vacuum is 
somewhat
complicated in a general curved space-time background, and different definitions
will in general lead to different results. Some particularly
important vacua are the asymptotic ones (assuming the space-time manifold to be
asymptotically flat), the adiabatic (assuming sufficiently small or slowly
varying curvature) and the conformal one. Other vacua, relying on the analogy
with thermal field theory, \cite{BD,Grib,Wald}, are the Unruh and Hartle-Hawking
ones, the applicability of these is usually restricted to Rindler and
Schwarzschild space-times, however. It turns out, by construction, that
our definition of a Green's function entails a natural vacuum. Recall the
spectral representation of the Green's function
\begin{displaymath}
    G(x,x') = \sum_\lambda \frac{1}{\lambda}\psi_\lambda^*(x')\psi_\lambda(x)
\end{displaymath}
In a second quantized formalism, these eigenfunctions are of course to be
replaced by operators, $\psi_\lambda(x) = \hat{\psi}_\lambda(x)|0_x\rangle$. 
Remembering that we're relying on comoving coordinates when determining $G$, 
we see that our vacuum
is that of a freely falling observer, and is thus well-defined on any manifold,
so there is no need to impose global constraints on its topology. The 
propagator, however, will vanish whenever $x$ and $x'$ are not causally 
related. Due to this
relationship with freely falling coordinates we will refer to our vacuum as the
freely falling one, or, somewhat more amusingly, the ``elevator vacuum.''\\
The question of when a candidate for a two point function is actually a
physically valid propagator, has been studied intensively by Bernard S. Kay and
coworkers, see \cite{Kay} for a review. They impose what they call the Hadamard 
condition, which, in
the heuristic formulation of DeWitt and Brehme \cite{DeWitt} (see also Wald 
\cite{Wald}), states that for 
$x'\approx x$
\begin{equation}
    G(x,x') \sim \frac{\sqrt{\Delta_{VVM}(x,x')}}{8\pi^2}\left(
    \frac{4}{\Delta(x,x')}+ v(x,x')\ln\frac{1}{2}\Delta(x,x')+w(x,x')\right)
\end{equation}
with $\Delta_{VVM}$ the Van Vleck-Morette determinant
\begin{displaymath}
    \Delta_{VVM}(x,x') \equiv -(g(x)g(x'))^{-1/2}\det(- \frac{1}{2}
        \nabla_\mu\nabla_\nu \Delta(x,x'))
\end{displaymath}
and where $v,w$ are smooth functions as $x'\rightarrow x$. Kay and Wald, 
\cite{Kay},
show that this is a physically reasonable requirement, and give a much more
rigorous formulation of it. We want to show that
our candidate, $G(x,x')$ (and hence also $D_m^n(x,x')$ and $S(x,x')$ for spin
$\neq 0$), meets this condition.\\
Now, writing $G(x,x')$ in the compact form
\begin{displaymath}
    G(x,x') = -2(4\pi)^{-2}\left[\alpha(x,x')K_1(z)+\beta(x,x')K_0(z)\right]
\end{displaymath}
with $z=\sqrt{\Delta(x,x')\tau_1(x,x')}$ and with $\alpha,\beta$ as given 
earlier in terms of $\Delta, \tau_n$, and using
\begin{eqnarray*}
    K_0(z) &=& -\ln\frac{1}{2}z+\psi(1)+O(z)\\
    K_1(z) &=& 2z^{-1}+O(1)
\end{eqnarray*}
we get
\begin{equation}
    G(x,x') = \tilde{\alpha}(x,x')\Delta^{-1}-\beta(x,x')\ln\frac{1}{2}\Delta
    +\gamma(x,x')
\end{equation}
where 
\begin{equation}
    -2(4\pi)^{-2}\alpha(x,x') = \tilde{\alpha}(x,x')\Delta^{-1/2}+
    \mbox{finite terms}
\end{equation}
and $\gamma(x,x')$ is well-behaved as $x\rightarrow x'$ as is $\tilde{\alpha}
(x,x')$ and $\beta(x,x')$. Our candidate thus has
the right form, and hence meets the Hadamard condition. Notice, however, that we
escape the logarithmic divergence as $\beta =O(\Delta)$, i.e. while we do have a
logarithmic term it does not lead to a divergence as $x'\rightarrow x$. This is
a most fortunate characteristic of our vacuum.\\
The way one usually finds a propagator, \cite{BD}, is by using the positive 
frequency solutions 
to the equation of motion -- the restrictions on the frequency being there 
to ensure
causal propagation. With the method put forward here, this requirement can be
relaxed as we no longer need to know a complete set of solutions. Causality, 
moreover,
is taken care of not by a process of time ordering, which might be difficult to
define rigorously in a general background over large distances, but in terms of
path ordering, or, if one wishes, in terms of the proper time of the comoving 
observer.
We thus get a causal propagator by restricting attention to time-like (for 
massive fields) or null geodesics (for massless ones).\\
We will make a few comments on the comparison of our propagator with that of
other authors, if such result exist (to the best of our knowledge), and it 
will turn
out, that the results presented here disagree with all other calculations, due
to the different notions of vacua. As our propagator does satisfy the Hadamard
condition it is physically valid. Moreover, it is more general and apparently
higher-order. It has for instance been argued that an adiabatic vacuum for
spatially flat Friedman-Robertson-Walker geometries is only a Hadamard state 
if it is taken to infinite adiabatic order, \cite{xx}.\\
Let us finally make a comment on the Schwinger-DeWitt expansion,
\begin{displaymath}
    G(x,x';\sigma) \sim (4\pi\sigma)^{-2}e^{-m^2\sigma-\frac{\Delta}{4\sigma}}
    \sum_{n=0}^\infty a_n(x,x')\sigma^n
\end{displaymath}
Our expansion can be written in a similar form, by simply Taylor expanding the
exponential. Denoting the coefficients by $b_n$, we have
\begin{eqnarray*}
    b_0 &=& 1\\
    b_1 &=& \tau_1\\
    b_2 &=& \frac{1}{2}\tau_1^2+\tau_2\\
    b_3 &=& \frac{1}{3!}\tau_1^3+2\tau_2\tau_1+\tau_3
\end{eqnarray*}
and so on, but $b_n\neq a_n$ in general, a difference stemming from the
different nature of the two expansions. At a first glance, our coefficients
seem to depend only on the curvature scalar and not on the Ricci or 
Riemann-Christoffel
tensors, but actually these tensors appear through the derivatives of the
geodesic distance squared. It turns out, as we have emphasized earlier, that it is
exactly this use of an exponential that makes a closed expression for the
expansion coefficients possible. An expression, moreover, valid for the general 
case $x\neq x'$ and reducing to a rather simple recursion relation for $x=x'$.
This latter case is important for practical calculations of, say, the
renormalized energy-momentum tensor, \cite{BD,Grib}), and with the usual
Schwinger-DeWitt expansion or with the modification due to Parker {\em et al.},
\cite{Parker}, only the first few can be found. On the contrary with the use of
the exponential expansion suggested here, one can in principle (and even in 
practice) systematically find the coefficients to whatever order one wishes.
We have used this possibility to calculate effective actions and energy-monentum
tensors elsewhere \cite{EffAct}, and further work is in progress on the 
applications of this expansion.\\
When studying coincidence limits, it is useful to evaluate
$\Delta$ for $x,x'$ infinitesmaly close. We thus concider two points $x,x'$ with
$x'_\mu=x_\mu+\epsilon t_\mu$ (in practice only one component of $t_\mu$
will be non-zero). We can then approximate the geodesic by a straight line with
a small correction. Letting, for a non-null geodesic, $\Sigma=t_\mu t^\mu$, 
as is common in the point splitting method, \cite{BD,Wald}, we get
\begin{eqnarray}
    \tau_1 &=& \frac{1}{2}(e^2+1)\tilde{f}_0(x)+\epsilon
    \left[(1-e^2-\sqrt{\Sigma})
    \tilde{f}_0(x)+\frac{1}{2}et^\mu\partial_\mu\tilde{f}_0(x)\right] +
    O(\epsilon^2)\\
    \tau_2 &=& \frac{1}{2}\Box_0 f_0(x) (1-4\epsilon\sqrt{\Sigma})+\frac{1}{32}
    (e^4-1)\epsilon t^\mu\Box_0\partial_\mu\Box_0 f_0(x)-\nonumber\\
    &&\frac{1}{4}\epsilon t^\mu\partial_\mu\Box_0 f_0(x)+O(\epsilon^2)\\
    \tau_3 &=& \frac{1}{144}\epsilon (e^6-1)t^\mu \Box_0\partial_\mu\Box_0 
    f_0(x)
    -\frac{1}{6}\epsilon t^\mu\partial_\mu(\partial_\nu f_0(x))^2-\nonumber\\
    && \frac{1}{3}(1-6\epsilon\sqrt{\Sigma})(\partial_\nu f_0(x))^2+
    O(\epsilon^2)
\end{eqnarray}

A quick application of the propagators found in this paper is the calculation
of mean fields. As we have just proven $G(x,x')$ is divergent as 
$x'\rightarrow x$, but we can use $\zeta$-function regularization to obtain a
finite value for $<\phi(x)^2>=G(x,x)$, the mean field. Going back to the 
expression of the propagator as an integral of the heat kernel, equation
(\ref{eq:Greens}), then we see that the $x'\rightarrow x$ limits gives a 
$\Gamma(-1)$-singularity. Such divergencies are to be removed by taking
principal values, \cite{princip}. We therefore get
\begin{equation}
    \langle \phi(x)^2\rangle_{\rm reg} = (4\pi)^{-2}\left[-(\gamma-1)\tau_1 +
    \tau_2\tau_1^{-1}+\tau_3\tau_1^{-2}+...\right] \label{eq:meanfield}
\end{equation}
as the final result.

\section{Examples}
To illustrate the technique developed above, we have chosen a few important
examples. The first is Rindler space-time, possibly the simplest non-trivial
example, and used for instance as an approximation to the Schwarzschild
solution. The second and third examples are probably the most important ones
from a purely theoretical point of view, as they are the classical solutions
of Friedman-Robertson-Walker and Schwarzschild respectively. We finish off 
with a
general conformally flat space-time (e.g. de Sitter space). For simplicity, we 
will just give the case of minimally coupled, massless scalar fields.
Furthermore, we will just list the simplest expressions, i.e. $\tau_1, {\cal A}$
and the coeffcicients $c_1,c_2,c_3,\tilde{\cal A},\tilde{\cal B},\tilde{\cal
C}$, the remaining terms (including all those for the fermions) follow by
suitable differentiations, which, although very simple to carry out, leads to
rather involved expressions.

\subsection{Rindler Space-Time}
The Rindler space-time is given by the line element
\begin{equation}
    ds^2 = (gz)^2dt^2-dx^2-dy^2-dz^2
\end{equation}
where $g$ is some constant, it is conformal to a wedge in Minkowski space-time
and represents an accelerated observer in flat space. 
With this the vierbeins can be chosen to be
\begin{equation}
    e_0^a = \left(\begin{array}{c}gz\\0\\0\\0\end{array}\right)\qquad
    e_1^a = \left(\begin{array}{c}0\\1\\0\\0\end{array}\right)\qquad
    e_2^a = \left(\begin{array}{c}0\\0\\1\\0\end{array}\right)\qquad
    e_3^a = \left(\begin{array}{c}0\\0\\0\\1\end{array}\right)
\end{equation}
A geodesic in this space-time takes the form 
\begin{eqnarray}
	t(\tau) &=& -\frac{t'-t}{\sinh a^{-2}} \sinh \frac{\tau}{a^2} +t\\
	x(\tau) &=& (x'-x)\tau+x\\
	y(\tau) &=& (y'-y)\tau+y\\
	z(\tau) &=& a\sqrt{\ln\cosh\frac{\tau}{a^2}+b}
\end{eqnarray} 
with $a,b$ two constants given by
\begin{equation}
	a\sqrt{b} = z\qquad\qquad a\sqrt{\ln\cosh a^{-2}+b}=z'
\end{equation}
Thereby
\begin{eqnarray}
    \sqrt{\Delta} &=&\int_0^1\left[g^2a^2(\ln\cosh\tau a^{-2}+b)\left(
	\frac{t'-t}{\sinh a^{-2}}\right)^2\cosh^2\tau a^{-2} -(x'-x)^2
	-(y'-y)^2\right.\nonumber\\
	&&\left. -\frac{1}{4}\tanh\frac{\tau}{a^{-2}} (\ln\cosh\frac{\tau}
	{a^2}+b)^{-1/2}\right]^{1/2}d\tau
\end{eqnarray}
which we have not been able to reduce any further. In practice, though, since 
the integration is over a compact interval, it is fairly easy to carry out
numerically.

\subsubsection{Scalar Field Propagator}
The quantities $f_0,\tilde{f}_0$, needed to find the scalar field propagator,
are seen to be (for a massless minimally 
coupled field -- masses and non-minially couplings can be accomodated by
adding $\xi R+m^2$ to these functions)
\begin{eqnarray}
    f_0 &=& -\frac{3}{4}z^{-2}\\
    \tilde{f}_0 &=& -\frac{7}{4}z^{-2}
\end{eqnarray}
From which we get the function $\tau_1$ to be equal to (remember $g_0=c_1=
\tilde{f}_0$)
\begin{equation}
	\tau_1(x,x') = -\frac{7}{4}e^{-2\sqrt{\Delta}} a^2\int_0^1 
	\frac{e^{2\tau}}{\ln\cosh\frac{\tau}{a^2}+b} d\tau
	-\frac{7}{8}(z^{-2}+z^{'-2})e^{-2\sqrt{\Delta}}
\end{equation}
As usual, the remaining coefficients $\tau_2,\tau_3,...$ are found by 
differentiation of this quantity in appropriate ways. 
The remaining $c_n$'s are found to be
\begin{equation}
    c_2 = -\frac{9}{4}z^{-4} \qquad c_3 = -\frac{1}{4}z^{-6}
\end{equation}
For a massless scalar field, an expression for the propagator is known already
\cite{Rindler}, namely
\begin{equation}
    G'(x,x') = \frac{\xi_4}{4\pi^2 zz'\sinh \xi_4(\xi_4^2-g^2(t-t')^2)}
\end{equation}
with
\begin{displaymath}
    \xi_4 \equiv {\rm Arccosh} \frac{z^2-z^{'2}+(x-x')^2+(y-y')^2}{2zz'}
\end{displaymath}
Defining coordinates
\begin{displaymath}
    x_0 = g^{-1}e^{g\xi}\sinh g\tau \qquad x_3 = g^{-1}e^{g\xi}\cosh g\tau
\end{displaymath}
the propagator can be rewritten as
\begin{displaymath}
    G'(x,x') = -\frac{1}{4\pi^2}\frac{g^2 e^{g\xi}}{\sinh \chi}
    \sum_{n=0}^\infty\frac{\chi}{g^2(\tau-\tau'+in\beta)^2-\chi^2-i\epsilon}
\end{displaymath}
where $\beta = \frac{2\pi}{g}$ is an inverse temperature (appearing due to
the Davies-Fulling-Hawking-Unruh effect, \cite{BD,Grib}) and
\begin{displaymath}
    \cosh\chi = \cosh g(\xi-\xi')+\frac{1}{2}g^2a^{-g(\xi+\xi')} ((x_1-x_1')^2+
    (x_2-x_2')^2)
\end{displaymath}
The integer $n$ is a kind of winding number.
For $x_2=x_2'=x_3=x_3'=0$ the geodesic distance becomes, \cite{Rindler}
\begin{displaymath}
    \Delta=4 g^{-2}\sinh^2(\frac{1}{2}g(\tau-\tau'))
\end{displaymath}
For simplicity we can set $x=x'=0, y=y'=0$ to get a
function of $t-t'$ and $z,z'$ only (notice that neither $G$, nor $G'$ are
functions of $z-z'$ but depends on more complicated combinations of $z,z'$). The
quantity $\xi_4$ is then simply
\begin{displaymath}
    \xi_4 = {\rm Arccosh}\frac{z^2-z^{'2}}{2zz'}
\end{displaymath}
Taylor expanding around $z=z',t=t'$ we get
\begin{eqnarray*}
    G'(x,x') &=& \frac{-1}{2\pi^3z^{'2}}+\frac{2g^2(t-t')^2}{\pi^5z^{'2}}+\\
    &&(z-z')\left(\frac{\pi-2}{2\pi^4z^{'3}}+\frac{2g^2(t-t')^2(6-\pi)}{\pi^6
    z^{'3}}\right)+\\
    &&(z-z')^2\left(\frac{6\pi-3\pi^2-8}{4\pi^5z^{'4}} +\frac{3g^2(t-t')^2
    (16-6\pi+\pi^2)}{\pi^7z^{'4}}\right)+O((z-z')^3,(t-t')^3)
\end{eqnarray*}
Since there is no singularity, this function does not satisfy the Hadamard
condition, and can thus at most be considered as an approximation to a physical
quantity.

\subsubsection{Gauge Boson Propagator}
In order to write down the propagator of a Yang-Mills field, we need to 
evaluate the matrices $\cal A,B,C$. As we do not have a tractable expression 
for the mean fields we cannot, however, write down these explictly; only 
their curvature (or abelian) contribution can be found explicitly. 
For the first of these matrices we then get
\begin{equation}
    \tilde{\cal A}_m^n = \frac{3}{4}\left(\begin{array}{cccc}
        z^{-2} & 0 & 0 & 0\\
        0      & 0 & 0 & 0\\
        0      & 0 & 0 & 0\\
        0      & 0 & 0 & z^{-2}
    \end{array}\right) +\mbox{ mean field terms}
\end{equation}
and similarly for the remaining matrices
\begin{eqnarray}
    \tilde{\cal B}_m^n &=& -\frac{9}{2}\left(\begin{array}{cccc} 
        z^{-4} & 0 & 0 & 0\\
        0      & 0 & 0 & 0\\
        0      & 0 & 0 & 0\\
        0      & 0 & 0 & z^{-4}
    \end{array}\right)+\mbox{ mean field terms}\\
    \tilde{\cal C}_m^n &=& \frac{9}{4}\left(\begin{array}{cccc}
        z^{-6} & 0 & 0 & 0\\
        0      & 0 & 0 & 0\\
        0      & 0 & 0 & 0\\
        0      & 0 & 0 & z^{-6}
    \end{array}\right)+\mbox{ mean field terms}
\end{eqnarray}
Thereby the coefficient $\cal A$ becomes
\begin{equation}
    {\cal A}_m^n = \frac{3}{7}\left(\begin{array}{cccc} 1 &0 &0 &0\\
    0 & 0 & 0 & 0\\ 0 & 0 & 0 & 0\\ 0 & 0 & 0 & 1\end{array}\right)
    \tau_1^{(s=0)}+\mbox{mean field terms}
\end{equation}
and quite similarly for $\cal B,C$
\begin{eqnarray}
    {\cal B}_m^n &=& \frac{3}{7}\left(\begin{array}{cccc} 
        1 & 0 & 0 & 0\\
        0 & 0 & 0 & 0\\
        0 & 0 & 0 & 0\\
        0 & 0 & 0 & 1\end{array}\right)\tau_2^{(s=0)}+\mbox{mean field terms}\\
    {\cal C}_m^n &=& \frac{9}{49}\left(\begin{array}{cccc} 
        1 & 0 & 0 & 0\\
        0 & 0 & 0 & 0\\
        0 & 0 & 0 & 0\\
        0 & 0 & 0 & 1\end{array}\right)\tau_3^{(s=0)}+\mbox{mean field terms}
\end{eqnarray}
Since these matrices are all diagonal (the only possibly non-diagonal term 
comes from $\langle gf_{b~c}^{~a}(\partial_n A^{mc}-\partial^m A^c_n)\rangle$, 
i.e. one 
of the two terms we cannot handle analytically), we can actually apply the 
Bessel functions to them without any great difficulty to simply obtain the
Bessel function of the (scalar) element multiplied by the constant $4\times 4$
matrix appearing in $\cal A,B,C$. Since the coefficient of proportionality
between the vector and scalar case is not the same for all three terms (it is
$3/7$ for the first two, and the square of this for the last) we do not get just
the scalar propagator multiplied by a constant $4\times 4$ matrix, instead we
get the coefficients $\alpha_1,\beta_1$ to be 
\begin{eqnarray}
    \alpha_1 &=&  2(4\pi)^{-2}\left(\sqrt{\frac{3}{7}}\left(
    \frac{\Delta}{4\tau_1}\right)
    ^{-1/2}+\sqrt{\frac{3}{7}}\left(\frac{\Delta}{4\tau_1}\right)^{1/2}\tau_2
    -\frac{3}{7}\frac{\Delta}{4\tau_1}\tau_3\right)\times \left(\begin{array}
    {cccc} 1 & 0 & 0 & 0\\ 0 & 0 & 0 & 0\\ 0 & 0 & 0 & 0\\ 0 & 0 & 0 & 1
    \end{array}\right)\nonumber\\
    &&\hspace{10mm}+\mbox{mean field terms}\\
    \beta_1 &=& (4\pi)^{-2}\frac{3}{7}\frac{\Delta}{2\tau_1}\tau_3\times 
    \left(\begin{array}{cccc} 
    1 & 0 & 0 & 0\\ 0 & 0 & 0 & 0\\ 0 & 0 & 0 & 0\\ 0 & 0 & 0 & 1
    \end{array}\right)
    +\mbox{mean field terms}
\end{eqnarray}
with $\tau_n$ given by the $s=0$ case.
For an abelian field such as the Maxwell field, the mean field terms become
irrelevant, and the expressions above becomes exact.

\subsection{Heat Kernel and Propagators in the Friedman-Robertson-Walker 
Metric} 
An important case is the cosmologically interesting Friedman-Robertson-Walker
space-time
\begin{equation}
    ds^2 = dt^2-a^2(t)(d\chi^2-f^2(\chi)(d\theta^2+\sin^2\theta d\phi^2))
\end{equation}
in which the vierbeins can be taken to read
\begin{equation}
    e_0^a = \left(\begin{array}{c}1\\0\\0\\0\end{array}\right)\qquad
    e_1^a = \left(\begin{array}{c}0\\a\\0\\0\end{array}\right)\qquad
    e_2^a = \left(\begin{array}{c}0\\0\\af\\0\end{array}\right)\qquad
    e_3^a = \left(\begin{array}{c}0\\0\\0\\af\sin\theta\end{array}\right) 
      \label{eq:RWvierb}
\end{equation}

\subsubsection{Scalar Field Propagator}
With the vierbeins given by equation (\ref{eq:RWvierb}) we get
\begin{eqnarray}
    f_0 &=&\frac{1}{4}\left(\frac{1}{e}e^m_\mu(\partial_m(ee^\mu_n))\right)^2-
    \frac{1}{2}\partial^n(\frac{1}{e}e^m_\mu\partial_m (ee^\mu_n))\nonumber\\
    &=& \frac{1}{4}a^{-2}(3\dot{a}-a^2f^{-1}f'-a^2\cot\theta)^2 -\frac{1}{2}
      (3\sin\theta
    \partial_t(a^{-1}\dot{a})-\dot{a}(f'\sin\theta+f\cos\theta))\nonumber\\
    &&+\frac{1}{2}a^{-2}(-3\dot{a}f^{-2}f'+a^2\partial_\chi(f^{-2}f'))+
    \frac{1}{2}\sin\theta\\
    \tilde{f}_0 &=& \frac{1}{4}a^{-2}(3\dot{a}-a^2f^{-1}f'-a^2\cot\theta)^2 
      +\frac{1}{2}(3\sin\theta
    \partial_t(a^{-1}\dot{a})-\dot{a}(f'\sin\theta+f\cos\theta))\nonumber\\
    &&-\frac{1}{2}a^{-2}(-3\dot{a}f^{-2}f'+a^2\partial_\chi(f^{-2}f'))-
    \frac{1}{2}\sin\theta
\end{eqnarray}
We note that $f_0$ and $\tilde{f}_0$ only differ in some relative signs. For 
the geodesic distance $\Delta$ we have not been able to find an analytical 
expression; in general one would then have to find this numericaly.\\
The coefficient $\tau_1$ becomes
\begin{eqnarray}
    \tau_1 &=& e^{-2\sqrt{\Delta}}\int_0^1
    e^{2\tau}\left[\frac{1}{4}a^{-2}\left(3\dot{a} - a^2 f^{-1}f'-a^2\cot\theta
    \right)^2+\frac{1}{2}(3\sin\theta \partial_t^2\ln
    a-\dot{a}(f'\sin\theta+f\cos\theta))\right.\nonumber\\
    &&-\left.\frac{1}{2}a^{-2}(-3\dot{a}f^{-2}f^{'2}+a^2\partial_\chi (f^{-2}f')
    -\frac{1}{2}\sin\theta\right]d\tau+
    \frac{1}{2}(\tilde{f}_0(x)+\tilde{f}_0(x'))e^{-2\sqrt{\Delta}}
\end{eqnarray}
For massless conformally coupled scalar fields some results are known
\cite{Dowker,BD,RW} in
the static case; for $K=-1$ one has
\begin{equation}
    G'_{\rm stat,-} (x,x') = \frac{\chi-\chi'}{4\pi^2 c\sinh(\chi-\chi')
    ((\chi-\chi')^2+(\eta-\eta'-i\epsilon)^2)}
\end{equation}
while for $K=+1$ (the Einstein universe) one has
\begin{equation}
    G'_{\rm stat,+}(x,x') = (8\pi^2 c(\cos(\eta-\eta'-i\epsilon)-\cos
    (\chi-\chi'))^{-1}
\end{equation}
with $\eta$ the conformal time, and $c$ given by $R=6K/c$. In order to compare
these results with ours, we must first notice that in the static case with
$\theta=\theta', \phi=\phi'$ the geodesics are trivial
\begin{displaymath}
    t(\tau)=a\tau+b\qquad\chi(\tau)=\alpha\tau+\beta
\end{displaymath}
with these simplifications, the coefficient $\tau_1$ becomes
\begin{displaymath}
    \tau_1 = e^{-2\sqrt{\Delta}}\int_0^1 e^{2\tau}\left[ \frac{1}{4} a_0^2
    (f^{-1}f'+\cot\theta)^2-\frac{1}{2}\partial_\chi(f^{-2}f')-\frac{1}{2}
    \sin\theta\right]d\tau +\tilde{f}_0e^{-2\sqrt{\Delta}}
\end{displaymath}
Now, since the geodesics are trivial, we can actually find the geodesic
distance
\begin{displaymath}
    \Delta = (t-t')^2-a_0^2(\chi-\chi')^2
\end{displaymath}
where $a_0$ is some constant (the value of the scale factor).\\ 
Taylor expanding the functions $G'_{{\rm stat},\pm}$ around $\eta=\eta',
\chi=\chi'$ taking $\theta=\theta',\phi=\phi'$ by symmetry, one finds
\begin{eqnarray*}
    G'_{\rm stat,-} &=& \frac{1}{4 c\pi^2(\eta-\eta')^2} -(\chi-\chi')
    \left(\frac{1}{4c\pi^2(\eta-\eta')^4}+\frac{1}{24c\pi^2(\eta-\eta')
    ^2}\right)\\
    &&+O((\eta-\eta')^3,(\chi-\chi')^3)\\
    G'_{\rm stat,+} &=& \frac{-1}{4c\pi^2(\eta-\eta')^2}-\frac{1}{48c\pi^2} -
    \frac{(\eta-\eta')^2}{160 c\pi^2}-\\
    &&(\chi-\chi')^2\left[\frac{1}{4c\pi^2(\eta-\eta')^4}+\frac{1}{24c\pi^2(\eta
    -\eta')^2}+\frac{11}{2880c\pi^2}+\frac{31(\eta-\eta')^2}{120960c\pi^2}
    \right]+\\
    &&O((\eta-\eta')^3,(\chi-\chi')^3)
\end{eqnarray*}
We notice that the singularity goes like $(\eta-\eta')^{-n}$, and is thus
independent of $\chi,\chi'$, we seem only to get a singularity when both
$\eta-\eta'$ and $\chi-\chi'$ tend to zero. On the other hand, Bunch and Davies,
\cite{RW}, have proven that the propagators obtained by a mode sum has the
following structure in the coincidence limit $x'\rightarrow x$ even in the
non-static limit.
\begin{displaymath}
    G' = \frac{-1}{16\pi^2\epsilon^2\Sigma}(1+O(\epsilon^2))
\end{displaymath}
and thus it seems to satisfy the Hadamard condition. It has been shown, however,
by Pirk, \cite{xx}, at least for spatially flat Friedman-Robertson-Walker 
space-times
that adiabatic vacua are Hadamard states if and only if they are of infinite
adiabatic order. Such subtleties are absent in our approach.\\ 
A very general form
for the propagator of a massive, conformally coupled scalar field in the flat
Friedmann-Robertson-Walker universe has been found by Charach and Parker, \cite{CP},
and it contains the older results of Narimi and Azuma, \cite{NA}, or Chitre and
Hartle, \cite{CH}, as special cases. This general formula is
\begin{displaymath}
	G(x,x') = \frac{i}{(2\pi)^2}\frac{\pi}{4tt'}\int e^{i{\bf k}\cdot({\bf x-x}')}
	H_{ik}^{(2)}(mt_>)\left[\frac{B_k^*}{C_k^*}e^{2\pi k}H_{ik}^{(2)}(mt_<)
	+H_{ik}^{(1)}(mt_<)\right]d^3k
\end{displaymath}
in terms of H\"{a}nkel functions. The coefficients $B_k,C_k$ are restricted by
$|C_k|^2e^{-k\pi}-|B_k|^2e^{k\pi}=\frac{\pi}{4}$, for various choices of these
one then gets the aforementioned results by Narimi and Azuma and by Banerjee and 
Hartle. In the above formula $t_>$ is the largest of $t,t'$, while $t_<$ is the
smallest of the two. That this actually satisfies the Hadamard condition in this
formulation is far from clear. Furthermore, the limit of $m\rightarrow 0$ seems
to be ill-defined as the H\"{a}enkel functions go like $z^{-\nu}$ where $z$ is
the argument and $\nu$ the order, contrary to this, our formula is valid for as well
massive as massless fields. Moreover, since the expression for $G(x,x')$ proposed in 
this paper does not involve an integral over the order of Bessel functions but only
integrals of geometric quantities along geodesics, it might be more convenient from a 
practical view point.\\
For finite temperature a result has been found by Banerjee and Mallik, 
\cite{BM}.

\subsubsection{The Gauge Boson Propagator}
In order to find the propagator of a spin 1 field we must first find the 
matrix $\tilde{\cal A}$ as defined above. Inserting the vierbeins 
(\ref{eq:RWvierb}) one can find
\begin{eqnarray}
    \tilde{\cal A}_n^m &=& \left(\begin{array}{cccc}
    3a^{-2}\dot{a}^2 & 2a^{-1}f^{-1}\dot{a}f' & a^{-1}\dot{a}\cot\theta & 0\\
    2a^{-1}f^{-1}\dot{a}f' & a^{-2}\dot{a}^2+2f^{-2}f^{'2} & f^{-1}f'\cot
      \theta & 0\\
    a^{-1}\dot{a}\cot\theta & f^{-1}f'\cot\theta & a^{-2}\dot{a}^2f^{-2}f^{'2}
      +\cot^2\theta & 0\\
    0 & 0 & 0 & a^{-2}\dot{a}^2f^{-2}f^{'2}+\cot^2\theta \end{array}\right)
      \nonumber\\
    &&\qquad+\mbox{ mean field dependent terms}
\end{eqnarray}
The matrix $\tilde{\cal B}$ is then $\Box_0=\eta^{mn}\partial_m\partial_n$ 
applied to this.
As we do not have a very tractable expression for the mean fields we will 
only list its curvature contribution together with that of $\tilde{\cal C}$ 
in table I.\\
The coefficient $\cal A$ becomes similarly
\begin{eqnarray}
    {\cal A}_n^m &=& e^{-2\sqrt{\Delta}}\int_0^1\left(\begin{array}{cccc}
    3\frac{\dot{a}^2}{a^2} & 2\frac{\dot{a}f'}{af} & \frac{\dot{a}}{a}
    \cot\theta & 0\\
    2\frac{\dot{a}f'}{af} & \frac{\dot{a}^2}{a^2}+2\frac{f^{'2}}{f^2} & 
    \frac{f'}{f}\cot\theta & 0\\
    \frac{\dot{a}}{a}\cot\theta & \frac{f'}{f}\cot\theta &
    \frac{\dot{a}^2f^{'2}}{a^2f^2}
      +\cot^2\theta & 0\\
    0 & 0 & 0 & \frac{\dot{a}^2f^{'2}}{a^2f^2}+\cot^2\theta \end{array}\right)
    e^{2\tau}d\tau\nonumber\\
    &&+\frac{1}{2}(\tilde{\cal A}_n^m(x)+\tilde{\cal A}_n^m(x'))
    e^{-2\sqrt{\Delta}}+\mbox{mean field terms}
\end{eqnarray}
Again, for a static Friedman-Robertson-Walker space-time this simplifies 
immensely to give
\begin{eqnarray*}
    {\cal A}_n^m &=& e^{-2\sqrt{\Delta}}\int_0^1\left(\begin{array}{cccc}
    3 & 0 & 0 & 0\\
    0 & 2f^{-2}f^{'2} & f^{-1}f'\cot\theta & 0\\
    0 & f^{-1}f'\cot\theta & \cot^2\theta & 0\\
    0 & 0 & 0 & \cot^2\theta \end{array}\right)e^{2\tau}d\tau+\\
    &&\frac{1}{2}(\tilde{\cal A}_n^m(x)+\tilde{\cal A}_n^m(x'))
    e^{-2\sqrt{\Delta}}
\end{eqnarray*}
with (as for the scalar case)
\begin{displaymath}
    \Delta = (t-t')^2 -a^2(\chi-\chi')^2
\end{displaymath}
Thus the Green's function for spin one contains essentially the same integrals
as for spin zero, albeit in a somewhat simpler configuration.

\subsection{Heat Kernel and Propagators in Schwarzschild Space}
The Schwarzschild metric is given by
\begin{equation}
    ds^2 = h(r)dt^2-\frac{1}{h(r)}dr^2-r^2d\Omega
\end{equation}
with
\begin{equation}
    h(r) = 1-\frac{2M}{r}
\end{equation}
where $M$ denotes the mass of the object generating the gravitational field.
Thus the vierbeins can be taken as
\begin{eqnarray}    
    e_0^a = \left(\begin{array}{c}\sqrt{h}\\0\\0\\0\end{array}\right)\qquad
    e_1^a = \left(\begin{array}{c}0\\h^{-1/2}\\0\\0\end{array}\right)\qquad
    e_2^a = \left(\begin{array}{c}0\\0\\r\\0\end{array}\right)\qquad
    e_3^a = \left(\begin{array}{c}0\\0\\0\\r\sin\theta\end{array}\right)
\end{eqnarray}
According to Chandrasekhar, \cite{Chandra}, the time-like geodesics are given by
\begin{eqnarray*}
    \frac{dt}{d\tau} &=& E/h(r)\\
    \left(\frac{dr}{d\tau}\right)^2 &=& E^2-h(r)(1+\frac{L^2}{r})\\
    \frac{d\theta}{d\tau} &=& 0\\
    \frac{d\phi}{d\tau} &=& \frac{L}{r^2}
\end{eqnarray*}
where $E$ is a constant (the energy) and where $L$ is the angular momentum. For
the radial case $L=0$, one finds
\begin{eqnarray*}
    t&=& E\sqrt{r_0^3}{2M}\left[\frac{1}{2}(\tau+\sin\tau)+(1-E^2)\tau\right]\\
      &&+2M\log\frac{\tan\frac{1}{2}\tau_H+\tan\frac{1}{2}\tau}{\tan\frac{1}{2}
      \tau_H-\tan\frac{1}{2}\tau}\\
    r&=& r_0\cos^2\frac{1}{2}\tau
\end{eqnarray*}
where
\begin{displaymath}
    r_0\equiv \frac{2M}{1-E^2}
\end{displaymath}
is the original position (the geodesics describe matter falling into the black
hole, or whatever is creating the gravitational field) and where
\begin{displaymath}
    \tau_H = 2{\rm Arcsin}~E
\end{displaymath}
Null geodesics, describing massless particles, are somewhat easier. The radial
solution is
\begin{eqnarray*}
    t &=& \pm r_*(\tau)+c_\pm\\
    r&=& \pm E\tau +c_\pm
\end{eqnarray*}
with
\begin{displaymath}
    r_*=r+2M\log\left(\frac{r}{2M}-1\right)
\end{displaymath}
and where $c_\pm$ are constants.

\subsubsection{Scalar Field Propagator}
For the scalar field we only need to evaluate the two quantities $f_0,
\tilde{f}_0$. With the vierbeins as given above, this is an easy task and we get
\begin{eqnarray}
     f_0 &=& r^{-3}(r-2M)+\frac{1}{4}M^2r^{-3}(r-2M)^{-1}+Mr^{-3}+
            \frac{3}{2}r^{-2}\nonumber\\
    && -Mr^{-4}(r-2M)^{-1}+\frac{1}{2}M^2r^{-4}(r-2M)^{-2}+M(r-2M)^{-1}\\
    \tilde{f}_0 &=& r^{-3}(r-2M)+\frac{1}{4}M^2r^{_3}(r-2M)^{-1}+Mr^{-3}-
            \frac{3}{2}r^{-2}\nonumber\\
    &&+Mr^{-4}(r-2M)^{-1}-\frac{1}{2}M^2r^{-4}(r-2M)^{-2}-M(r-2M)^{-1}
\end{eqnarray}
which can then be inserted into the definition of $\tau_1$
\begin{displaymath}
    \tau_1 = e^{-2\sqrt{\Delta}}\int_0^1\tilde{f}_0e^{2\tau}d\tau+\frac{1}{2}
    (\tilde{f}_0(r)+\tilde{f}_0(r'))e^{-2\sqrt{\Delta}}
\end{displaymath}
for a radial null-geodesic, the integral over $\tau$ can actually be carried out
and we get
\begin{eqnarray*}
    \int_0^1\tilde{f}_0(r(\tau))e^{2\tau}d\tau &=& (cE)^{-2}(c(E-2M)-EM) +
    (E(c+E))^{-2}e^2(E^2+c(E-2M)-3EM)+\\
    &&2E^{-3}e^{-2cE^{-1}}(E-2M)({\rm Ei}(2+2cE^{-1})-{\rm Ei}(2cE^{-1}))+
    \frac{1}{4}M^2\times\\
    &&\left[\frac{1}{2}c(E-c-2M)-\frac{1}{4}E(E-2M)-2M^2+\frac{1}{2}e^2c(c+E+2M)
    \right.+\\
    &&\frac{1}{4}e^2E(E+2M)+2M^2e^2+8E^{-1}M^3e^{-2(c-2M)/E}\times\\
    &&\left.({\rm Ei}(2(c+E-2M)/E) -{\rm Ei}(2(c-2M)/E))\right]+\\
    &&M\left[\frac{3}{2}c^{-1}E^{-2}-\frac{1}{2}e^2E^{-2}(2c+3E)(c+E)^{-2}
    +2E^{-3}e^{-2c/E}\times\right.\\
    &&\left.({\rm Ei}(2c/E)-{\rm Ei}(2+2c/E))\right]-\\
    &&\frac{3}{2}\left[(cE)^{-1}-e^2E^{-1}(c+E)^{-1}+2e^{-2c/E}E^{-2}(
    {\rm Ei}(2+2c/E)-{\rm Ei}(2c/E))\right]+\\
    &&M\left[-\frac{1}{24 c^3E^3M^3}\left(3c^2E(E+2M)+3cE^2M+4M^2
    (E^2+2c^2+cE)\right)\right.\\
    &&e^2(24E^3M^3(c+E)^3)^{-1}\left(3c^2E(E+2E^2+2M)+3E^4+E^3M(9+15c)+\right.\\
    &&\left.4M^2(4E^2+2c^2+5cE)\right)+(12M^3E^4)^{-1}e^{-2c/E}(3E^2+6EM +8M^2)
    \times\\
    &&\left({\rm Ei}(2c/E)-{\rm Ei}(2+2c/E)\right)+\\
    &&(16EM^4)^{-1}e^{-2c/E}\left({\rm Ei}(2c/E)-{\rm Ei}(2+2c/E)+\right.\\
    &&\left.\left.{\rm Ei}(2(c+E-2M)/E)e^{-4M/E}-{\rm Ei}(2(c-2M)/E)e^{-4M/E}
    \right)\right]-\\
    &&\frac{1}{2}M^2\times\left[(12EM^2c^3)^{-1}-(16E(2M-c)M^4)^{-1}
    +\right.\\
    &&e^2(16E(2M-E-c)M^4)^{-1}-e^2(12EM^2(c+E)^3)^{-1}+\\
    &&\frac{1}{8}(EM^5)^{-1}e^{-2c/E}\left({\rm Ei}(2+2c/E)
    -{\rm Ei}(2c/E)\right)+\\
    &&(48c^2E^3M^4)^{-1}\left(c(9E^2+12EM+8M^2)+6E^2M+4EM^2\right)+\\
    &&e^2(48E^3(c+E)^2M^4)^{-1}\left(9E^2(c+E+2M)+12cEM+4M^2(3E+2c)\right)+\\
    &&(48E^4M^4)^{-1}e^{-2c/E}\left({\rm Ei}(2c/E)-{\rm Ei}(2+2c/E)\right) +\\
    &&\left.
    (8E^2M^5)^{-1}e^{-2(c-2M)/E}\left({\rm Ei}(2(c+E-2M)/E)-{\rm Ei}
    (2(c-2M)/E)\right)\right]-\\
    &&ME^{-1}e^{-2(c-2M)/E}\left({\rm Ei}(2(c+E-2M)/E)-{\rm Ei}(2(c-2M)/E)
    \right]
\end{eqnarray*}
which looks very complicated but is actually just a combination of polynomials,
exponentials and exponential integral functions.\\
Unfortunately, the geodesic distance $\Delta$ is rather difficult to find 
analytically in Schwarzschild
space-time, only radial geodesics are easily obtained. In many aplications, 
though, these actually suffice. 
Along these geodesics the differential operators appearing in the expressions
for $\tau_2,\tau_3$ are
\begin{eqnarray}
    \partial_p &=& e_p^\mu\dot{x}_\mu\frac{\partial}{\partial\tau} =
    (h^{-1/2}\dot{t} -h^{1/2}\dot{r})\frac{\partial}{\partial\tau}\\
    \Box_0 &=& \eta^{pq}\partial_p\partial_q = (h^{-1}\dot{t}\ddot{t}-h\dot{r}
    \ddot{r})\frac{\partial}{\partial\tau}+(h^{-1}\dot{t}^2-h\dot{r}^2)
    \frac{\partial^2}{\partial\tau^2}
\end{eqnarray}
Fairly simple expressions. But we will not write down the explicit form for
$\tau_2,\tau_3$ which follows from the application of these operators as the
results are too involved.\\
Candelas and Jensen, \cite{Candelas}, have also found a propagator for the
Schwarzschild background. They use the Hartle-Hawking vacuum (i.e. the Green's
functions is demanded to vanish as the spatial distance tends to infinity and to
be periodic in imaginary time), with this they arrive at
\begin{eqnarray*}
    G(-i\tau,r,\theta,\phi,-i\tau',r',\theta',\phi') &=& \frac{i}{32\pi^2 M^2}
    \left[\sum_{l=0}^\infty (2l+1)P_l(\cos\gamma) P_l(\xi_<)Q_l(\xi_>)+\right.\\
    &&\left.\sum_{n=1}^\infty \frac{1}{n}\cos(n\kappa(\tau-\tau'))
    \sum_{l=0}^\infty(2l+1)P_l(\cos\gamma)p_l^n(\xi_<)q_l^n(\xi_>)\right]
\end{eqnarray*}
where $P_l,Q_l$ are Legendre polynomials, $p_l^n,q_l^n$ are solutions to 
\begin{displaymath}
    \left[\frac{d}{d\xi}(\xi^2-1)\frac{d}{d\xi}-l(l+1)-\frac{n^2(\xi+1)^4}{ 16
    (\xi^2-1)}\right] f=0
\end{displaymath}
where $\xi=\frac{r}{M}-1$ and
\begin{eqnarray*}
    \xi_< &=& \min(\xi,\xi')\\
    \xi_> &=& \max(\xi,\xi')\\
    \cos\gamma &=& \cos\theta\cos\theta'+\sin\theta\sin\theta'\cos(\phi-\phi')
\end{eqnarray*}
They prove that
\begin{displaymath}
    \lim_{x'\rightarrow x}G(x,x') = \frac{i}{8\pi^2\Delta(x,x')}
    +\frac{i}{12(8\pi M)^2}\frac{1-\left(\frac{2M}{r}\right)^4}{1-\frac{2M}{r}}
    +\frac{iF(r)}{(8\pi M)^2}
\end{displaymath}
with $F(r)$ small. Like our results, then, Candelas and Jensen finds a Green's
function with only the $\Delta^{-1}$-singularity, whether they also get a
$\Delta \log\Delta$ term is, however, not so clear, since not much is known
about $q_l^n,p_l^n$ and the sums involving these and/or Legendre polynomials is
difficult to actually carry out. If one calculate the finite part of our
propagator as $x'\rightarrow x$, one finds
\begin{eqnarray*}
   &&-\frac{M}{32\pi^2}\left(48 M^3-M^2 r(48+52M^2)+Mr^2(19+159M^2)-158M^2 r^3
   +\right.\\
    && \left.66Mr^4-r^5(10-4M^2)-2Mr^6\right)(8r^7(r-2M)^3)^{-1}+\\
    &&\frac{1}{32\pi^2}\left(48 M^3-M^2 r(48+52M^2)+Mr^2(19+159M^2)-158M^2 r^3
   +\right.\\
    &&\left.66Mr^4-r^5(10-4M^2)-2Mr^6\right)^2(24r^{10}(r-2M)^6)^{-1}-\\
    &&\frac{1}{32\pi^2}\left(240M^4 -M^3r(384+216M^2)+M^2r^2(198+666M^2)
    -Mr^3(30+793M^2)+\right.\\
    &&\left. 461M^2r^4-132Mr^5+r^6(15-4M^2)+2Mr^7\right)(8r^7(r-2M)^3)^{-1}
\end{eqnarray*}
i.e. one finds more terms which diverge as one approaches the Schwarzschild
radius. A discrepancy likely to be due to (1) the different vacua (we do not
have an explicit temperature) and (2) the different orders of the respective
approximations.
 
\subsubsection{Gauge Boson Propagator}
The curvature contribution to the matrix $\tilde{\cal A}$ is in this case, 
with the vierbeins as given above, easy to find and we get
\begin{eqnarray}
    \tilde{\cal A}_m^n &=& \frac{1}{2}\left(\begin{array}{cccc}
    {\scriptsize h^2(\partial_rh^{-1/2})^2 } & 0 & 0 & 0\\
    0 & {\scriptsize h^2(\partial_rh^{-1/2})^2+2\frac{h}{r^2}} & {\scriptsize 
        \frac{h^{1/2}\cot \theta}{r^2}} & 0\\
    0 & {\scriptsize \frac{h^{1/2}\cot \theta}{r^2}} & {\scriptsize 
        \frac{h+\cot \theta}{r^2} } & 0\\
    0 & 0 & 0 & {\scriptsize \frac{h+\cot \theta}{r^2}} \end{array}\right) 
    \nonumber\\
    &&\qquad+\mbox{ mean field terms}
\end{eqnarray}
whereby the matrices $\tilde{\cal B},\tilde{\cal C}$ become as follows
\begin{eqnarray}
    \tilde{\cal B}_m^n &=& -\frac{1}{4}\left(\begin{array}{cc}
    \frac{2M^2(27M^2 - 32Mr + 10r^2)}{r^6(r-2M)^2} & 0 \\
    0 & \frac{2(243M^4 - 400M^3r + 240M^2r^2 - 62Mr^3 + 6r^4}{r^6(r-2M)^2}\\
    0 & \frac{(2M - r)(10M - 4r - 10M\cos 2\theta + 3r\cos 2\theta)
    \cot\theta\csc^2\theta}{\sqrt{1 - \frac{2M}{r}}r^6}\\ 
    0 & 0\end{array}\right.\nonumber\\
    &&\nonumber\\
    &&\left.\begin{array}{cc}    
    \frac{(2M - r)(10M - 4r - 10M\cos 2\theta + 3r\cos 2\theta)\cot\theta
    \csc^2\theta}{\sqrt{1 - \frac{2M}{r}}r^6} & 0\\ 
    \frac{54 M^2 - 38 M r+6r^2}{r^6}-\frac{(14 M-6r)\cot 2\theta}{r^4}+
    \frac{\cot\theta\csc^2\theta}{r^4} & 0\\
    0 & \frac{2r^2-2M^2-2Mr}{r^2} -\frac{(14M-6r)\cot\theta}{r^5}
    +\frac{\cot\theta\csc^2\theta}{r^4}\end{array}\right)\nonumber\\
    &&\qquad\qquad+\mbox{ mean field terms}
\end{eqnarray}
where we, for reasons of space, have split-up the matrix $\tilde{\cal B}$ and
\begin{eqnarray}
    \tilde{\cal C}_m^n &\approx &
    \frac{1}{12}\times\nonumber\\
    &&\hspace{-30mm}\left(\begin{array}{cccc}
    \frac{4M^4(2r-3M)^2}{r^7(r-2M)^5} & 0 & 0 & 0\\
    0 & \frac{4(27M^3 - 34M^2r + 14Mr^2 - 2r^3)^2}{r^7(r-2M)^5} & \frac{(5M -
    2r)^2\cot^2\theta}{r^6(r-2M)^2}+\frac{(r-2M)\csc^4\theta}{r^7} & 0\\
    0 & \frac{(5M-2r)^2\cot^2\theta}{r^6(r-2M)^2}+
    \frac{(r-2M)\csc^4\theta}{r^7} &
    \frac{6M-2r}{r^4}-\frac{4\cot^2\theta}{r^5(r-2M)}+\frac{\csc^4\theta}{r^6} &
    0\\
    0 & 0 & 0 &\frac{6M-2r}{r^4}-\frac{4\cot^2\theta}{r^5(r-2M)
    }+\frac{\csc^4\theta}{r^6} \end{array}\right)\nonumber\\
    &&\qquad\qquad+\mbox{ mean field terms}
\end{eqnarray}
And this can then be inserted into the formula (\ref{eq:YMprop}) for the 
propagator $D_m^n(x,x')$.

\subsection{Conformally Flat Space-Times}
As our last example we will consider a conformally flat space-time with line 
element
\begin{equation}
    ds^2 = C(\eta)^2(d\eta^2-\delta_{ij}dx^idx^j)
\end{equation}
This could for instance be de Sitter space or a general
spatially flat Friedman-Robertson-Walker metric in arbitrary dimensions. For 
Friedman-Robertson-Walker
space-time the conformal time $\eta$ is defined to be $\int^ta^{-1}dt$ and 
$C(\eta)$
is just the usual scale factor $a(t)$ expressed in this new coordinate, 
whereas for de Sitter space-time $\eta=-\alpha\exp(-t/\alpha)$ where $\alpha = 
\sqrt{12/R}$ with $R$ the curvature scalar (i.e. a constant in this case). 
The calculations
will be carried out in $d=4$ dimensions, but can of course easily be 
generalized to any $d$.\\
For the metric as given above, we can choose the vierbeins to be
\begin{equation}
    e_0^a =\left(\begin{array}{c}C(\eta)\\ 0\\ 0\\ 0\end{array}\right)
    \qquad e_1^a=\left(\begin{array}{c}0\\ C(\eta)\\ 0\\ 0\end{array}\right)
    \qquad e_2^a=\left(\begin{array}{c}0\\ 0\\ C(\eta)\\ 0\end{array}\right)
    \qquad e_3^a=\left(\begin{array}{c}0\\ 0\\ 0\\ C(\eta)\end{array}\right)
\end{equation}

\subsubsection{Scalar Boson Case}
The d'Alembertian can be written as
\begin{equation}
    \Box = \Box_0 + {\cal E}_a\partial^a = \Box_0 + 3\frac{\dot{C}}{C^3}
    \partial_\eta
\end{equation}
where $\dot{C}=\partial_\eta C$.
We then have
\begin{eqnarray}
    f_0 &=& \frac{21}{4}\dot{C}^2C^{-4}-\frac{3}{2}\ddot{C} C^{-3}\\
    \tilde{f}_0 &=&\frac{33}{4}\dot{C}^2C^{-4}+\frac{3}{2}\ddot{C}C^{-3}
\end{eqnarray}
whereby we get
\begin{eqnarray}
    \tau_1 &=& e^{-2\sqrt{\Delta}}\int_0^1\left(\frac{33}{4}\dot{C}^2C^{-4}
    +\frac{3}{2}\ddot{C}C^{-3}\right)e^{2\tau}d\tau\nonumber\\
    &&+\frac{1}{4}\left(\frac{33}{2}\dot{C}(\eta)^2C(\eta)^{-4}+\frac{33}{2}
    \dot{C}(\eta')^2C(\eta')^{-4}+3\ddot{C}(\eta)C(\eta)^{-3} + 3\ddot{C}(\eta')
    C(\eta')^{-3}\right)e^{-2\sqrt{\Delta}}\nonumber\\
\end{eqnarray}
and the coefficients $c_2, c_3$ are found to be
\begin{eqnarray}
    c_2 &=&-\frac{3}{4}(C^{-5}C^{(4)}+C^{-6}(C+1)\dot{C}\dot{\ddot{C}})
    -3C^{-6}\ddot{C}^2+\frac{1}{4}C^{-7}(51 C+174)\dot{C}^2\ddot{C}\nonumber\\
    &&-21C^{-7}\dot{C}^4-\frac{105}{2}C^{-8}\dot{C}^4\\
    c_3 &=& \frac{147}{4}C^{-10}\dot{C}^2\ddot{C}^2 - 147
    C^{-11}\dot{C}^4\ddot{C}+147 C^{-12}\dot{C}^{-6}
\end{eqnarray}
It turns out that in the particular case of de Sitter space-time, another 
formula for the propagator of a scalar field is known in the literature 
\cite{BD,deSit}, namely\footnote{For thermal fields, a propagator was also found by
Chardury and Mallik, \cite{CM}, but it will not be discussed here.} 
\begin{equation}
    G'(x,x') = \frac{1}{16\pi\alpha^2}(\frac{1}{4}-\nu^2)\sec \pi\nu
    F(\frac{3}{2}+\nu,\frac{3}{2}-\nu;2;1+\frac{(\eta-\eta'-i\epsilon)^2-(
      {\bf x}-{\bf x}')^2}
    {4\eta\eta'}) \label{eq:BD}
\end{equation}
with
\begin{displaymath}
    \nu^2\equiv \frac{9}{4}-12(m^2 R^{-1}+\xi)
\end{displaymath}
Clearly this can not be the final answer as it diverges for a massless minimally
coupled field (in this case $\nu=\frac{3}{2}$).
Let us nonetheless compare it with our result obtained by the method 
described in this paper. The first thing we notice is that, even on the
level of series expansion as in equation (\ref{eq:Bessel}), our expression 
is manifestly divergent as $x\rightarrow x'$ (due to the
$\Delta^{-1}, \log\Delta$ terms), whereas the expression 
(\ref{eq:BD}) only diverges when all the terms in the series
expansion of the hypergeometric function is included (it then goes to 
$\Gamma(-1)$). More explicitly, by letting $x\rightarrow x'$ along a non-null
geodesic with tangent $t^\mu$, $t^\mu t_\mu \equiv \Sigma = \pm 1$, one has
\cite{deSit} (for $d=2$, but the form is the same in any dimension, only the
coefficients change)
\begin{eqnarray*}
    G(x,x') &\approx & -(4\pi)^{-1}\left[ 2\gamma +\ln \frac{1}{2}\epsilon^2
    \Sigma R + \psi(\frac{1}{2}+\mu)+\psi(\frac{1}{2}-\mu)+\right.\\
    && \frac{1}{6}\epsilon^2\Sigma R-\epsilon^2\Sigma(\xi R+m^2)\left(2\gamma+
    \ln \frac{1}{2}\epsilon^2 \Sigma R+\right.\\
    &&\left.\left. \psi(\frac{3}{2}+\mu)+\psi(\frac{3}{2}-\mu)-2\right)\right]
\end{eqnarray*}
with 
\begin{displaymath}
    \mu = \sqrt{\frac{1}{4}-2\xi-m^2\alpha^2}
\end{displaymath}
Now, in this case $\Delta = \frac{1}{2}\Sigma \epsilon^2$, so what we capture
here is only the $\log\Delta$ divergence and not the more common $\Delta^{-1}$
one, consequently it does not satisfy the Hadamard condition.
The result (\ref{eq:BD}) follows from a mode summation using an adiabatic 
vacuum, the failure to meet the Hadamard condition is then likely to be a result
of not going to infinite adiabatic order. The adiabatic vacuum
has the advantage of being very similar to flat space-time, but is, alas, 
only possible in some space-times. The Green's function
following from the heat kernel as put forward here, on the other hand, comes 
with a ``natural vacuum'', which apparently  
works in all cases. The difference between our result and (\ref{eq:BD}) can 
thus come from many sources:
\begin{itemize}
    \item Ambiguity in the definition of a vacuum (\cite{BD} uses adiabatic 
      vacuum which does not always work).
    \item Ambiguity in the definition of a propagator (the Green's function 
      as defined in this paper is rather unique, and from this all
    propagators -- positive frequency, advanced, retarded, Whightmann etc. -- 
      can be found).
    \item The approximations made (our result is highly non-pertubative and 
      seems to include more terms than in (\ref{eq:BD}).
\end{itemize}
This also summarizes the differences found in the other specific geometries.

\subsubsection{Gauge Boson Case}
The tensor 
\begin{displaymath}
    {\cal E}_n^{mp} \equiv \eta^{mq}(\partial_ne_q^\mu-\partial_q e_n^\mu) 
    e_\mu^p
\end{displaymath}
is found to have the following non-vanishing components only
\begin{equation}
    {\cal E}_0^{ii} = {\cal E}_i^{0i} = -\dot{C}C^{-1}
\end{equation}
with no summation over the index $i=1,2,3$ implied. From this we get the
mean-field independent part of $\tilde{\cal A}$ to be
\begin{equation}
    \tilde{\cal A}_n^m = -\frac{9}{4}\delta^m_n\dot{C}^2C^{-2} +\mbox{mean field
    terms}
\end{equation}
whereby
\begin{eqnarray}
    {\cal A}_n^m &=& -\frac{9}{4}\delta_n^me^{-2\sqrt{\Delta}}\int_0^1
    \left(\frac{d}{d\eta}\ln C\right)^2e^{2\tau}d\tau \nonumber\\
    &&-\frac{9}{16}\delta_n^me^{-2\sqrt{\Delta}}\left(\dot{C}(\eta)^2
      C(\eta)^{-2}
    +\dot{C}(\eta')^2C(\eta')^{-2}\right)+\mbox{mean fields}
\end{eqnarray}
For the matrix-coefficients $\tilde{B},\tilde{C}$ we get similarly
\begin{eqnarray}
    \tilde{\cal B}_n^m &=& \frac{9}{8}\delta_n^m\left(2C^{-4}\ddot{C}^2
    +2C^{-4}\dot{C}\dot{\ddot{C}}-10 C^{-5}\dot{C}^2\ddot{C} +6 C^{-6}\dot{C}^4
    +2\dot{C}^2\ddot{C}C^{-4}-2C^{-5}\dot{C}^4\right)\nonumber\\
    &&\qquad+\mbox{  mean field terms}\\
    \tilde{\cal C}_n^m &=& \frac{81}{12}\delta_n^m\left(\dot{C}\ddot{C} C^{-2} -
    C^{-3}\dot{C}^3\right)^2+\mbox{mean field terms}
\end{eqnarray}
And once more one can get the remaining terms by simple differentiation of 
these.

\section{Conclusion and Outlook}
We have developed a way to determine the propagators of quantum fields
in curved space-time explicitly. It is, for practical
reasons, presently only possible to calculate an approximation to 
the propagator. This approximation, while excellent for scalar and spinor fields
and relatively good for spin 1 gauge bosons is, in the latter case, 
particularly for non-abelian gauge fields, rather 
complicated to work with so one might hope to change this by altering the 
approach
by which we found this expression. In all the cases, though, the practical 
calculation of the Green's function,
was reduced to that of calculating a few functions (for spin zero and one half,
$f_0,\tilde{f}_0$) or three matrices 
(for vector bosons, $\tilde{\cal A},\tilde{\cal B},\tilde{\cal C}$) upto 
interactions with other fields than the gravitational background. \\
It turned out, however, that our non-perturbative propagators, though satisfying
the Hadamard condition and having a very simple physical meaning, differed from
what other authors had found in various cases. This difference can be traced
back to the uniqueness of the vacuum as presented here, and which differs, in
general, from the adiabatic or conformal vacuum usualy used. The ambiguity in
the definition of a vacuum is related to the insufficiency of the naive particle
concept in curved space-times. By always referring to a local comoving
observer, we seem to have arrived at a better defined particle concept, since we
are then able to relate to a flat space-time at each point along the trajectory,
and of course, in flat space-time there is no problem with the particle
concept, as the general success of ordinary quantum field theory shows. It
should furthermore be emphasized that the method put forward in this paper is
very general, as we do not to assume any particular feature of the metric. If
the space-time under consideration does show some particular features (such as
weak or slowly varying curvature, asymptotic or conformal flatness and so on),
then of course, the calculation of the propagator by our method can be
simplified, or rather adjusted, by then only including the necessary terms in
the general expansion of the heat kernel. As these coefficients are actually
rather easy to find this adjustment is straightforward in each particular 
case.\\
Looking at our results for, say, Friedman-Robertson-Walker and 
Schwarzschild geometries,
we notice the appearance of terms which survive if the flat space limit is
understood as $a(t)\rightarrow\infty$ and $M\rightarrow 0$ respectively. These
effect are most likely due to the difference in topologies; for 
Friedman-Robertson-Walker
space-times we still have $K=\pm 1,0$ even in the limit $a\rightarrow \infty$,
and in the Schwarzschild case, the limit $M\rightarrow 0$ corresponds to
Minkowski space with the origin removed. It is not at present clear what the
physical implications of such corrections are, however. When doing perturbation
theory in flat space-time, one should of course in principle take into account
that we do not live in Minkowski space, and thus use, say, the $a\rightarrow
\infty$ limit of a Friedman-Robertson-Walker propagator. One would expect any
such effect to be very small, but it is not {\em a priori} certain as the
results presented here are in deed non-perturbative, and even though the
perturbative effect is very small, the true, non-perturbative effect need not
be. Given the present day limitations on accelerator power, the only chance of
seeing a quantum gravitational effect, or an effect coming from the coupling of
the quantum fields to the space-time geometry, is either by studying the early
universe and/or black holes or to look for some non-perturbative effect.\\
Naturally, the method of calculating the heat kernel as presented here in this
paper is in no way restricted to quantum field theory in curved space-time. It
should also be of considerable mathematical interest, as it is known that the
heat kernel of the Laplace-Beltrami operator contains information about the 
large scale structure of the underlying manifold. Non-perturbative quantum field
theory in ordinary Minkowski space is of course just a particular example, and
the uses there include quark-gluon plasma.\\
The uses for the propagators include refinements in determining the spectrum 
of Hawking emission from black holes as well as the emission of gravitational 
waves from the space-time singularity (the Big Bang), applications that we hope 
to entertain in subsequent papers. Another application, which we have not yet 
considered, is the evaluation of the trace-anomaly of the energy-momentum 
tensor in general. Lastly we should note that, having obtained expressions for
the propagators in this paper and for the one-loop effective actions in a
previous one, we are now in a position to find the effective action to any loop
order (at least formally). A paper on this is in progress.

%-------- bibliography-------------------------

%------- tables ----------------
\begin{table}[htb]
\hspace{-10mm}
\begin{tabular}{|c|c|c|}\hline
$(m,n)$ & ${\cal B}_n^m$ & ${\cal C}_n^m$ \\ \hline
$(0,0)$ & $\scriptsize 18\left(\frac{\dot{a}}{a}\right)^4-30\frac{\dot{a}^2
\ddot{a}}{a^3}+2\left(\frac{\ddot{a}}{a}\right)^2
           +2\frac{\dot{a}\ddot{a}}{a^2}$ & $\scriptsize \frac{9}{4}\left(
\frac{\dot{a}\ddot{a}}{a^2}-\frac{\dot{a}}{a}\right)^2
                +\frac{1}{4}a^{-2}\left(\frac{\ddot{a}}{a}-\left(\frac{\dot{a}}
{a}\right)^2\right)^2
                \left[\left(\frac{f'}{f}\right)^2+f^{-2}\cot^2\theta\right]$\\ 
        &     & $\scriptsize +\frac{1}{4}\dot{a}^2a^{-6}\left(\frac{f''}{f}-
\left(\frac{f'}{f}\right)^2\right)+\dot{a}^2
                a^{-6}f^{-2}\sin^{-4}\theta$\\ \hline
$(0,1)$ & $\scriptsize 2\frac{f'}{f}\left(2\left(\frac{\dot{a}}{a}\right)^3-
           3\frac{\dot{a}\ddot{a}}{a^2}+\frac{a^{(3)}}{a}\right)$ &
             $\scriptsize \left(\frac{\dot{a}\ddot{a}}{a^2}-\left(\frac{\dot{a}
}{a}\right)^2\right)
              \left(\frac{\ddot{a}}{a}-\left(\frac{\dot{a}}{a}\right)^2\right)
\frac{f'}{f}$\\
        & $\scriptsize -2\frac{\dot{a}}{a^3}\left(2\left(\frac{f'}{f}\right)^3-
3\frac{f'f''}{f^2}+\frac{f^{'''}}{f}\right)$ 
              & $\scriptsize \frac{1}{2}\left(\frac{f''}{f}-\left(\frac{f'}{f}
\right)^2\right)\left(\frac{f'f''}{f^2}
                -\left(\frac{f'}{f}\right)^3\right)\frac{\dot{a}}{a^3}+
\frac{\dot{a}f'}{a^3f^2\sin^4\theta}$\\ \hline
$(0,2)$ & $\scriptsize \cot\theta\left(2\left(\frac{\dot{a}}{a}\right)^3-3
\frac{\dot{a}\ddot{a}}{a^2}+\frac{a^{(3)}}{a}
            \right)$ & $\scriptsize \left(\frac{\dot{a}\ddot{a}}{a^2}-\left(
\frac{\dot{a}}{a}
                   \right)^3\right)\left(\frac{\ddot{a}}{a}-\frac{\dot{a}}{a}
\right)\cot\theta$\\
        & $\scriptsize -2\frac{\dot{a}\cos\theta}{a^3f^2\sin^3\theta}$ & 
$\scriptsize 
                \frac{1}{4}\frac{\dot{a}}{a^3}\left(\frac{f''}{f}-\left(
\frac{f'}{f}\right)^2\right)\cot\theta
                +2\frac{\dot{a}}{a^3}\frac{\cos\theta}{\sin^5\theta}f^{-2}$  
\\ \hline
$(1,1)$ & $\scriptsize 6\left(\frac{\dot{a}}{a}\right)^4-10\frac{\dot{a}^2
\ddot{a}}{a^3}+2\left(\frac{\ddot{a}}{a}\right)^2 
            +2\frac{\dot{a}\ddot{a}}{a^2}$ & $\scriptsize \frac{1}{4}\left(
\frac{\ddot{a}}{a}-\left(\frac{\dot{a}}{a}\right)^2
          \right)\left(\frac{f'}{f}\right)^2+\frac{1}{4}\left(\frac{\dot{a}
\ddot{a}}{a^2}-\left(\frac{\dot{a}}{a}\right)^3\right)$\\
        & $\scriptsize 12a^{-2}\left(\frac{f'}{f}\right)^4-20a^{-2}\frac{f'f''}
{f^3}$ &
                $\scriptsize +\frac{1}{4}\left(\frac{f''}{f}-\left(\frac{f'}{f}
\right)^2\right)^2\left[\left(\frac{\dot{a}}{a}
                    \right)^2+\cot^2\theta\right]a^{-2}$\\ 
        & $\scriptsize +2a^{-2}\left(\frac{f''}{f}\right)^2+2a^{-2}\frac{f'f''}
{f^2}$ & $\scriptsize \left(\frac{f'f''}{f^2}-
            \left(\frac{f'}{f}\right)^3\right)^2a^{-2}+\left(\frac{f'}{f^2}
\right)^2a^{-2}\sin^{-4}\theta$\\ \hline
$(1,2)$ & $\scriptsize a^{-2}\cot\theta\left(2\left(\frac{f'}{f}\right)^3-3
\frac{f'f''}{f^2}+\frac{f^{'''}}{f}\right)$ & 
                $\scriptsize \frac{1}{4}\cot\theta
                \frac{f'}{f}\left(\frac{\ddot{a}}{a}-\left(\frac{\dot{a}}{a}
\right)^2\right)^2$\\ 
        & $\scriptsize -2\frac{\cos\theta}{\sin^3\theta} a^{-2}\frac{f'}{f^3}$ 
& $\scriptsize \frac{3}{4}\left(\frac{f'f''}{f^2} -
                \left(\frac{f'}{f}\right)^3\right)\left(\frac{f''}{f}-\left(
\frac{f'}{f}\right)^2\right)a^{-2}\cot\theta$ \\ 
        &     &  $\scriptsize 2a^{-2}\frac{f'\cos\theta}{f^3\sin^5\theta}$\\ 
\hline
$(2,2)$ & $\scriptsize 9\left(\frac{\dot{a}}{a}\right)^4-10\frac{\dot{a}^2
\ddot{a}}{a^3}+2\left(\frac{\ddot{a}}{a}\right)^2
            +2\frac{\dot{a}\ddot{a}}{a^2}$ & $\scriptsize \frac{1}{4}\left(
\frac{\dot{a}\ddot{a}}{a^2}-\left(\frac{\dot{a}}{a}
            \right)^3\right)^2+\frac{1}{4}\left(\frac{\ddot{a}}{a}-\left(
\frac{\dot{a}}{a}\right)^2\right)^2\cot^2\theta$\\
        & $\scriptsize +6a^{-2}\left(\frac{f'}{f}\right)^4-10a^{-2}
\frac{f'f''}{f^3}+2 a^{-2}\left(\frac{f''}{f}\right)^2$ &
                    $\scriptsize +\frac{1}{4}\left(\frac{f''}{f}-\left(
\frac{f'}{f}\right)^2\right)^2a^{-2}\cot^2\theta+\frac{1}{4}
                    \left(\frac{f''f'}{f^2}-\left(\frac{f'}{f}\right)^3
\right)^2a^{-2}$\\
        & $\scriptsize +2a^{-2}\frac{f'f''}{f^2}-2a^{-2}f^{-2}\frac{1+2\cos^2
\theta}{\sin^4\theta}$ & 
             $\scriptsize +\left(\left(\frac{\dot{a}}{a}\right)^2+\left(
\frac{f'}{f}\right)^2+4\cot^2\theta\right)a^{-2}f^{-2}
            \sin^{-4}\theta$\\ \hline
$(3,3)$ & as $(2,2)$ & $\scriptsize \frac{1}{4}\left(\frac{\dot{a}\ddot{a}}
{a^2}-\left(\frac{\dot{a}}{a}\right)^3\right)^2+
                \frac{1}{4}\left(\frac{f'f''}{f^2}-\left(\frac{f'}{f}\right)^3
\right)^2a^{-2}$\\ 
        &      & $\scriptsize \frac{1}{8}a^{-2}f^{-2}\frac{\cos^2\theta}
{\sin^6\theta}$\\ \hline
\end{tabular}
\caption{The mean field independent contribution to the symmetric matrices 
$\cal B$ and $\cal C$ for a Friedman-Robertson-Walker space-time.}
\end{table}


\begin{thebibliography}{99}
\bibitem{BD} N.D. Birrel, P.C.W. Davies: Quantum Fields in Curved Space, 
Cambridge 1982.
\bibitem{Grib} A.A. Grib, Mamayev, V.M. Mostepanenko: Vacuum Quantum Effects 
in Strong Fields, Friedmann Lab. Publishing, St.Petersburg 1994.
\bibitem{Wald} R.M. Wald: Quantum Field Theory in Curved Spacetime and Black 
Hole Thermodynamics, University of Chicago Press, 1994.
\bibitem{Kay} B.S. Kay, R.M. Wald: Phys.Rep. {\bf 207} (1991)49 and references 
therein.
\bibitem{Casimir} K.Bormann, F.Antonsen: ``Casimir Effect in Curved Space 
(formal 
developments)'', Proceedings of the Third International Alexander Friedmann 
Seminar, St.Petersburg 1995.
\bibitem{EffAct} F.Antonsen, K.Bormann: ``Effective Actions For Spins $0,
\frac{1}{2},1$ in Curved Space-Times.'' (submitted).
\bibitem{Parker} L. Parker, D.J. Toms: Phys.Rev.{\bf D31}(1985) 953; 
Phys.Rev.{\bf D31}(1985) 2424; I. Jack, L. Parker: Phys.Rev.{\bf D31}(1985) 
2439.
\bibitem{FT} L.H. Ford, D.J. Toms: Phys.Rev.{\bf D25} (1982) 1510.
\bibitem{path} J.D. Bekenstein, L. Parker: Phys.Rev.{\bf D23} (1981) 2850.
\bibitem{geod} D.E. Neville: Phys.Rev.{\bf D28} (1983) 265.
\bibitem{GR} I.S. Gradstheyn, I.M. Ryzhik: Table of Integrals, Series and
Products, Academic Press, New York 1967.
\bibitem{DeWitt} B.S. DeWitt, R. Brehme: Ann.Phys.(NY) {\bf 9} (1960)220.
\bibitem{Rindler} W. Troost, H. Van Dam: Nucl.Phys. {\bf B152} (1979)442; B. 
Linet: gr-qc/9505033; R.Emparan: Phys.Rev. {\bf D51} (1995)5716.
\bibitem{Chandra} S. Chandrasekhar: The Mathematical Theory of Black Holes, 
Clarendon Press, Oxford 1983.
\bibitem{RW} T.S. Bunch, P.C.W. Davies: Proc.Roy.Soc. London {\bf A357} (1977)
381.
\bibitem{CP} Ch. Charach, L. Parker: Phys.Rev.{\bf D28}(1981) 3023.
\bibitem{NA} H. Narimi, T. Azuma: Prog.Theor.Phys. {\bf 59}(1978) 1538.
\bibitem{CH} B.M. Chitre, J.B. Hartle: Phys.Rev.{\bf D16}(1977) 251.
\bibitem{BM} N. Banerjee, S. Mallik: Phys.Rev.{\bf D44}(1991) 3770.
\bibitem{Candelas} P.Candelas, B.P. Jensen: Phys.Rev.{\bf D33} (1986)1596.
\bibitem{Dowker} J.S. Dowker, R. Critchley: Phys.Rev.{\bf D15} (1977)1484.
\bibitem{deSit} T.S. Bunch, P.C.W Davies: Proc.Roy.Soc. London {\bf A360} 
(1978) 117.
\bibitem{CM} A. Chardury, S. Mallik: Phys.Rev.{\bf D36}(1987) 1259.
\bibitem{xx} K.-T. Pirk: Phys.Rev. {\bf D48} (1993)3779.
\bibitem{princip} S.K. Blau, M. Visser, A. Wipf: Nucl.Phys.{\bf B310} (1988)163.
\end{thebibliography}
\end{document}